# Measuring the concept of PID literacy: user perceptions and understanding of persistent identifiers in support of open scholarly infrastructure

George Macgregor[1, 2] (0000-0002-8482-3973), Barbara S. Lancho-Barrantes[3] (0000-0001-9994-8886) & Diane Rasmussen Pennington[2, 4] (0000-0003-1275-7054)

[1]Scholarly Publications & Research Data, Scholarly Research Communications, University of Strathclyde, Glasgow (UK)

[2] Department of Computer & Information Sciences, University of Strathclyde, Glasgow (UK)

[3] School of Architecture, Technology and Engineering, University of Brighton, Brighton (UK)

[4] School of Computing, Edinburgh Napier University, Edinburgh (UK)

## Abstract

The increasing centrality of persistent identifiers (PIDs) to scholarly ecosystems and the contribution they can make to the burgeoning 'PID graph' has the potential to transform scholarship. Despite their importance as originators of PID data, little is known about researchers' awareness and understanding of PIDs, or their efficacy in using them. In this article we report on the results of an online interactive test designed to elicit exploratory data about researcher awareness and understanding of PIDs. This instrument was designed to explore recognition of PIDs (e.g. DOIs, ORCIDs, etc.) and the extent to which researchers correctly apply PIDs within digital scholarly ecosystems, as well as measure researchers' perceptions of PIDs. Our results reveal irregular patterns of PID understanding and certainty across all participants, though statistically significant disciplinary and academic job role differences were observed in some instances. Uncertainty and confusion were found to exist in relation to dominant schemes such as ORCID and DOIs, even when contextualized within real-world examples. We also show researchers' perceptions of PIDs to be generally positive but that disciplinary differences can be noted, as well as higher levels of aversion to PIDs in specific use cases and negative perceptions where PIDs are measured on an 'activity' semantic dimension. This work therefore contributes to our understanding of scholars' 'PID literacy' and should inform those designing PID-centric scholarly infrastructures, that a significant need for training and outreach to active researchers remains necessary.

This is a peer reviewed, accepted author manuscript of the following research article: Macgregor, G., Lancho-Barrantes, B. S., & Rasmussen Pennington, D. (Accepted/In press – 21 February 2023). Measuring the concept of PID literacy: user perceptions and understanding of persistent identifiers in support of open scholarly infrastructure. *Open Information Science*.

# Introduction

Persistent identifiers (PIDs), such as Digital Object Identifiers (DOIs) or the Open Researcher and Contributor ID (ORCID), are becoming central to the operation of scholarly systems, especially open scholarly infrastructure. PIDs enable persistent reference to nodes on the web — people, places, and things — facilitating the generation of vast scholarly 'PID graphs' capable of mapping connections between different entities within scholarly landscapes (Cousijn et al., 2021). CERN note that PIDs make these entities 'discoverable', by identifying them uniquely and reliably; 'accessible', by successfully resolving them even if the entity changes its location; 'useable', by referencing a specific version or state of an entity; 'interoperable', by expressing trust through transparency and provenance; and, finally, 'assessable', by contributing to an interconnected network of entities (CERN, 2020). These properties present opportunities for a variety of innovations in the scholarly communications space, not least in the identification, discovery, and retrieval of scholarly entities (Ananthakrishnan et al., 2020). The transformational potential of PIDs in future open research ecosystems is such that a growing number of research organizations now insist upon their use wherever possible (Bosman et al., 2021).

The adoption and use of PIDs by scholarly systems is predominantly a concern for the operational administrators of these systems, and they are typically features with which many scholarly users have only passive interaction. However, many users are also important *contributors* of PID data comprising the PID graph and are frequently the origin of this PID data (Dappert et al., 2017). Such contributions are made through the creation, supply, and interlinking of heterogeneous textual objects, datasets, software, research instruments, equipment and the related PIDs these items may generate, such as for people, organizations, or other abstract entities. If the PID graph is to demonstrate reliable growth and adequate relational depth, it will be necessary for scholarly contributors to participate meaningfully with PID centric systems and to demonstrate a level of 'PID literacy' in their (re)use and creation. Despite their importance as originators of PID data, little is known about scholars' awareness and understanding of PIDs, or their efficacy in using them. A knowledge gap also exists in our understanding of academics' perceptions of PIDs and the extent to which their introduction to scholarly infrastructure is perceived positively.

In this paper we report on the results of an online interactive test designed to elicit exploratory data about the awareness and understanding of PIDs among scholars. This test instrument was designed to explore scholars' recognition of PIDs and the extent to which scholars correctly apply PIDs within digital scholarly ecosystems, thereby measuring their 'PID literacy' — a concept we introduce in this article. The instrument also sought to measure scholars' perceptions of PIDs. Data elicitation in the study of perception in information science is often qualitative in nature (Cox & Abbott, 2021; Purvis et al., 2017). However, using aspects of Osgood's original measurement approaches (Osgood, 1957), the instrument instead deployed semantic differential measurement techniques to better capture metrics on the 'semantic dimensions' of scholars' perceptions of PIDs. The relative novelty of our approach is the measurement of scholars' perceptions using these semantic dimensions and the measurement of perceptual distances across participant groups.

While this exploratory work is motivated from a data perspective and its implications for the PID graph, it also provides a useful snapshot of scholarly users' digital scholarship competencies and therefore also contributes to wider body of information literacy knowledge. It also provides useful perceptual insights into scholars' thinking around the increased use of PIDs in scholarly ecosystems. Taken together this greatly informs future technical approaches to the implementation of PIDs but also informs the educational and training requirements necessary to support scholars in their (re)use of PIDs.

The paper will first provide explanatory background on persistent identification, PIDs, and the PID graph before going on to review related research and theory. A formal statement of the research motivation and research questions to be addressed then follows. We continue by describing the research instrument design, followed by results and a discussion of those results.

## Persistent identification and the PID graph

A persistent identifier (PIDs) provides a unique and *persistent* reference to an entity which is normally accessible over the Internet (McMurry et al., 2017). Such PIDs provide long-term identification for these entities but also actionability by being encoded as a Uniform Resource Identifier (URI) (e.g. Weigel, Kindermann, and Lautenschlager 2014). A PID registry service — usually designed to assign PID URI prefixes, register new PIDs, and provide the necessary technical infrastructure manage PIDs — will also typically provide metadata describing the entity, contextualizing its potential (re)use. One common form of PID which has come to typify the identification of scholarly resources on the web is the digital object identifier (DOI). The notion underpinning a DOI is that it remains static ('persistent') over time and, when dereferenced on the web, will always resolve to the identified resource, irrespective of whether its digital location may have changed in the intervening period (International DOI Foundation, 2017). This general PID approach has since been adopted within digital scholarly infrastructure to persistently identify a wide variety of people, places and things on the web, from researchers (e.g. ORCID[1]) to research projects (e.g. Research Activity iD (RAiD)[2]) to cell lines (e.g. Research Resource ID (RRID)[3]), some reusing existing Handle.net infrastructure or devising different but similar service mechanisms to enable the assignation, registration, and resolution of PIDs (Cousijn et al., 2021). Figure 1 provides an example for the following Research Organization Registry (ROR) PID, the European Organization for Nuclear Research (CERN). ROR is used to persistently identify research related organizational entities and, in this case, identifies CERN: https://ror.org/01ggx4157. In the same way that DOI registries gather metadata about the entities to which they are assigning DOIs, so associated structured data for this PID is gathered by the ROR registry, available via the ROR API: https://api.ror.org/organizations/01ggx4157.

*<Take in Figure 1>*

Although the concept of source identification in scholarship is not new (Kaplan, 1965), PID approaches under discussion in this paper have emerged within the context of digital object management where the issue of 'link rot' and persistence in the scholarly record have become important considerations in maintaining a healthy scholarly publications and research ecosystem (Jones et al., 2016). PIDs enable the improved citation and tracking of scholarly entities because their identifiers — whether of people, places, or things — remain persistent over time. This mitigates problems arising from link or reference rot and enables continued re-use of whatever is identified by the PID, either by user or machine. For this reason, PIDs are increasingly considered an important insurance mechanism within scholarship, assuring that scholarly verification, reproducibility, and replicability remain possible within an age of evolving digital scholarly infrastructure (Ivie and Thain 2018; Meadows, Haak, and Brown 2019). Additionally, the improvements in tracking made possible by PIDs offer scientometric potential, with services like Crossref becoming valuable datasets in better measuring research impact, knowledge growth, and scholarly communications trends (Hendricks et al., 2020).

It should be noted, however, that PIDs in themselves do not and cannot guarantee persistence. For example, investigations by Klein and Balakireva (2022) of DOIs — arguably the most ubiquitous PID type — suggests widespread DOI request failures and inconsistent machine responses from organizations using them. Members of the same research team have also proposed their Memento 'Robust Links' approach as a means of improving the reliability of Uniform Resource Locator (URL) and URI-based referencing on the web, including with respect to PIDs (Klein et al., 2018). PIDs are therefore only persistent insofar as a PID registration service commits to resolving them, or insofar as a publisher commits to updating a PID registry with the current location of a web resource. Meanwhile there are explorations of alternative PID models for services like open repositories, where the use repository Open Archive Identifier (OAI) identifiers — a unique identifier for a metadata record held within a repository or OAI compliant software platform — offers a decentralized mechanism for contributing to scholarly data graphs (Knoth et al., 2022). Despite these caveats and developments, PIDs can and do foster greater persistence of scholarly entities and help to unambiguously identify them, thereby encouraging resource discovery and reuse.

---

[1] Open Researcher and Contributor ID (ORCID): https://orcid.org/
[2] Research Activity Identifier (RAiD): https://www.raid.org.au/
[3] Research Resource Identifiers (RRID): https://www.rrids.org/

*<Take in Figure 2>*

The ability to unambiguously identify entities — combined with greater certainty of persistence — has enabled the encoding of relational associations between PIDs and their associated metadata, in turn generating complex but valuable scholarly graph-based networks (Um et al., 2020). By extending many of the conventions established by the Resource Description Framework (RDF) (Macgregor, 2009) and its use within Linked Open Data (LOD) in particular (Treloar, 2011), the so-called 'PID graph' presents opportunities for improved resource discovery, inferencing / reasoning, research aggregation, and so forth by establishing connections between different entities within the research landscape. Research Activity IDs (RAiD), for instance, can persistently identify research activities such as an ongoing research project. This RAiD can then be referenced within scholarly article metadata thereby creating a relational association and the colocation of articles arising from the same research project. But the RAiD itself can also be used to reference associated entities within its 'metadata envelope', as in Figure 2. In this diagrammatic example, the RAiD PID references DOIs of research datasets associated with the research activity, an identifier from the Global Research Identifier Database (GRID) and the International Standard Name Identifier (ISNI) scheme, a Group Identifier (GiD), and even another related RAiD, referenced as a sub-project (Janke et al., 2017).

Numerous scholarly graphs have emerged in recent years but not all generate their entity relations entirely using PIDs and instead use knowledge extraction and inferencing techniques, or combinations thereof (Atzori et al., 2018; Manghi et al., 2020; Schirrwagen et al., 2020). Though the PID graph is dependent upon metadata that demonstrates high levels of expressiveness, the graph is simpler because graph construction, maintenance, and relational associations are based entirely on the PIDs themselves. This aids graph scaling as the computational overhead associated with knowledge extraction, inferencing, and de-duplication does not exist (Cousijn et al., 2021). Suffice to state that important contributors to the PID graph include — but is not limited to — DataCite,

CrossRef, ORCID, ROR, ISNI, FundRef, RRID, and RAiD. The nature of this paper is such that a fuller treatment of the PID graph exceeds our scope; however, useful technical background is provided by many others in the literature (Aquino et al., 2017; Cousijn et al., 2021; Dappert et al., 2017; Klump & Huber, 2017; Meadows et al., 2019).

It is also apposite to note that emerging models of scientific communication entail even greater levels of PID specificity. So-called 'nanopublication' models (Bucur, Kuhn, and Ceolin 2020) and emerging platforms such as Octopus.ac[4] seek to disaggregate the components of research into verifiable, citable chunks. Such models disaggregate research into persistently identifiable components, such as research problem, hypothesis, method, analysis and so forth (e.g. (Freeman 2022)); however, within other models disaggregation can extend to specific paragraphs within a scientific paper (Bucur et al. 2020). These models are beyond the indicative cases to be noted later in this paper but nevertheless point to an emerging trend which is seeking to expose scholars to the 'PID-ification' of almost all aspects of their research.

## User research and introducing 'PID literacy'

An extensive body of PID literature has emerged since the early 2000s, especially within digital library, digital repository, and scholarly infrastructure research, but focuses almost exclusively on PID infrastructure, technical governance, PID types, policies, metadata profiles, and so forth (e.g. Chandrakar, 2006; Foulonneau & André, 2008; Koehler, 1999; Nelson & Allen, 2002; Simons & Richardson, 2013; Weigel et al., 2013). This literature has been critical to the technical evolution of PID-based technologies but does not address the socio-technical side of PID growth. For example, little is known about scholars' understanding and perception of PIDs, or the extent of users' 'PID literacy' — a concept which we will introduce in this section.

PIDs are largely concerned with citing entities. Their broader efficacy is dependent upon their accurate reproduction by others. Since our direct understanding of scholars' interaction with PIDs is minimal, it is useful to consider tangential areas of research, especially work exploring scholars' capacity for identifying scholarly entities, such as their citation habits and competency. Although the competencies required to use and understand PIDs is higher, the routine of citing sources within academic work is somewhat cognate since it requires that sources — i.e. entities — be accurately identified (Kaplan, 1965). Such work most notably investigates scholars' citation habits and can inform our understanding of the present research context (Cano, 1989). For example, observations by Eugene Garfield (Garfield, 1974, 1990) in the early phases of bibliometrics found frequent and numerous citation errors in scholarly articles, introduced by authors during drafting. Misidentification of journals, misspelling of author names, and name mis-orderings were found to be common. Some errors have been found to be associated with a lack of citation verification (e.g. authors copying faulty citations from a faulty source) (Broadus, 1983), a phenomenon which persists and has been termed 'referencing misbehaviour' (Liang et al., 2014). There are indications too of variation in the citing quality and behaviour of scholars across subjects, with patterns noted as being peculiar to specific subjects, compounding the problem (dos Santos et al., 2022).

---

[4] Octopus.ac: https://www.octopus.ac/

Early studies also found inaccuracies within 40% of all sampled citations (Key & Roland, 1977), with subsequent work by Asano et al. finding up to 48% of all articles containing one or more errors, of which author names and article titles were found to be the source of 70% of errors (Asano et al., 1995). Interestingly, a recent longitudinal study investigating citation accuracy in articles published at ten year intervals in a well-known journal title (between 1991 and 2019) reported an overall citation error rate of 40% (Logan, 2022), with little fluctuation in this error rate between decade intervals. This body of work suggests that in many cases scholars' ability to accurately identify and cite scholarly sources is not as good as it should be, despite a perceived improvement in digital scholarship ability among scholars.

Citation is as old as scholarship itself — but fluency with PIDs includes an additional understanding of web technology and the notion that URIs can reference abstract entities on the web. The information science sub-domain of information literacy has for some time noted the general challenges arising from the growth in digital academia (Behrens, 1994; Bruce, 1995; Sorapure et al., 1998). Findings from studies produced in previous decades have alerted academia to information literacy deficiencies present within some academic disciplines (Boon et al., 2007; Secker, 2004). Influential information literacy conceptual models, most notably the 'Framework for Information Literacy for Higher Education' proposed by The Association of College & Research Libraries (ACRL), has subsequently recognized potential deficiencies within scholarly communities (ACRL, 2016). The ACRL Framework presents 'Scholarship as Conversation' as a core component of information literacy, within which scholars should develop "familiarity with the sources of evidence, methods, and modes of discourse" used to engage in 'scholarly conversations'.

More recently conceptions of information literacy have broadened to 'digital literacy', encompassing notions of web literacy and technical skills sets (Alexander et al., 2017), or the more expansive overarching concept of 'metaliteracy' (Mackey & Jacobson, 2017). These broader digital competencies have been severely tested in recent years as scholars have been forced — some with great difficulty — to engage in online teaching and digital research as a consequence of the Covid-19 pandemic (Heriyanto et al., 2022). Suffice to state that recent work studying digital literacy competencies suggest that such skill deficiencies are common among some academic groups (Ong, 2021), with deficiencies identified to be found to be across scholarship more generally. For example, a literature review by Basilotta-Gómez-Pablos et al. analysed 56 articles that studied the digital competencies of university teaching staff, finding that 'low or medium–low' digital competence was dominant (Basilotta-Gómez-Pablos et al., 2022).

Though increasingly unreflective of current information literacy habits given its publication age, Wouters and de Vries (2004) found wide variability in the way in which web-hosted scholarly resources were referenced using hyperlinks by scholars, noting a lack of standardization in hyperlink conventions. Notwithstanding that these problems have a long history in academia, it could be speculated that more recent neglect among doctoral student cohorts is inherited from an assumption that younger scholars are 'digital natives' and therefore more likely to be digitally competent or information literate, a notion that has been widely refuted by recent evidence (Judd, 2018). Indeed, information literacy experiments conducted by Greer and McCann (2018) involving the citation behaviour of final year undergraduate students observed that supposedly 'digitally literate' students "do not understand URLs", with many unable to distinguish between official and unofficial URLs or redirects, or even correctly identify a digital source on the web (Greer & McCann, 2018). It is therefore reasonable to assume that these behaviours may be carried forward to doctoral level, and perhaps beyond; and an inability to even understand URLs does raise questions about the potential extent of PID competencies in academia more generally.

The notion of 'PID literacy' is not a concept which has been defined in the literature. We introduce it here to help us conceptualize the expected competencies a typical scholar might need in order to interact with PIDs effectively. It is possible to identify some typical scenarios which help us understand how PID literacy might manifest itself in users. Indicative cases where PID literacy might be exercised could include one or more of the following scenarios:

- Correctly supplying an ORCID for all contributing authors when submitting an article via a manuscript submission system.

- Providing PIDs to related datasets within an article manuscript (e.g. a data availability statement); but could equally apply to providing PIDs to related software, research instruments, samples, etc.

- Correctly reproducing and referencing funders, grants, and other abstract entities using PIDs.

- Curating research related PIDs (e.g. funders, organizations, collaborators, etc.) within a RAiD metadata 'envelope' (a location for descriptive information about research activities to organized within RAiD).

- Correctly identifying and citing (by PID) different expressions of the same scholarly work, as expressed through preprints, accepted author manuscripts, versions of records, etc.

- Understanding when creation, use or reuse of a PID might be relevant and knowledge of suitable or applicable PID types.

The above noted examples are merely indicative and any number of alternative PID related scenarios could be imagined. Nevertheless, on this basis, we could propose that a PID literate scholar might display the following competencies:

1. An understanding of persistent identification in scholarship, when it should be used, and its importance to the scholarly record and the wider PID graph.

2. An ability to accurately identify, reproduce, and cite PIDs in scholarship activities.

3. Cognizance of adjacent PID types relevant to scholars' community of practice, such as those devised to identify scholarly 'things' other than academic papers.

PID literacy, as conceived here, is not a holistic conception of information literacy but instead could be considered an extension or sub-literacy of existing frameworks, most notably ACRL's 'Scholarship as Conversation' frame (ACRL, 2016). This 'frame', and its companion document on 'Research Competencies in Writing and Literature' (ACRL, 2021), highlight the expected 'knowledge practices' and 'dispositions' of scholars if they are to meaningfully interact with research and scholarly communications processes. These practices and dispositions clarify the expected competencies of scholars but stop short of capturing PID literacy behaviours.

## Motivation and research questions

We noted in earlier sections that, despite being largely ignored in the discussions surrounding PID development, scholarly users are frequently an important source of data to the PID graph, whether they are aware they are the source of this data or not. Even perfunctory contributions greatly enrich the PID graph, furthering the relational and inferential potential of the graph — and are essential to the creation and interlinking of heterogeneous entities, including different 'expressions' of textual objects, datasets, software, research instruments, equipment and the related PIDs these items may in turn generate. Better understanding the nature of scholars' understanding and perceptions of PIDs therefore motivates our research. It motivates our work, not only because of the contributions that scholars may or may not make to the PID graph because of their PID literacy; but because scholars' literacy with PIDs within wider scholarly infrastructures will increasingly be exercised.

The research questions that emerge from our motivation are therefore as follows:

- **RQ1 – *PID familiarity***: To what extent are users familiar with the notion of the persistent identification of scholarly entities, their purpose, and understanding of the URIs that underpin them?

- **RQ2 – *Identifying identifiers***: To what extent can users correctly identify and distinguish between common PID types and their purpose?

- **RQ3 – *PID perceptions***: How are PIDs perceived by users and what levels of disciplinary or job role differences can be observed across academic groups, if any — and can any specific PID literacy gaps be identified?

- **RQ4 – *Habits***: Current state of PID (re)use; what are the habits? Do they routinely contribute (knowingly or unknowingly) to the PID graph? Are they habitually (re)using them (e.g. citation/references, in the creation and/or linking of scholarly entities, etc.)?

## Methods

## Research instrument design

### *Overview*

An online research instrument was designed as the principal data collection method. This instrument included aspects of an online interactive test designed to elicit experimental data about scholars' awareness, perceptions, and understanding of PIDs. For simplicity of administration, the instrument was delivered using online questionnaire technology but designed to test participants' capacity for PID recognition and PID perception. The research instrument is openly available as a .qsf file or .pdf, along with research data arising from this study (Macgregor et al., 2022).

The instrument was divided into four distinct sections: 1) computer self-efficacy, 2) PID recognition tests, 3) PID perception measurement, and 4) PID (re)use habits. The details of each section are explained in the sections below.

Participants were asked to complete a total of 33 tasks (including some questions in section 4), producing a combination of descriptive, nominal, and ordinal data. Owing to the exploratory nature of this research, some tasks were multifaceted, particularly in section 3 where semantic differential measurement techniques were deployed, resulting in 13 bipolar adjectives per PID concept (of which there were four). Others, such as section 2, attempted to simulate as best as possible real-world PID challenges by using screenshots taken from active scholarly infrastructure. A brief additional section at the end of the instrument captured basic demographic data, (e.g. country of origin, academic job role, discipline — as defined by the All Science Journal Classifications (ASJC) scheme) (BARTOC, 2021)). Participant response data were anonymized.

### *Instrument sections*

Since we are using quite a diverse population of scholarly participants, there is the possibility that computer efficacy in some of these sub groupings may differ significantly, thereby influencing their ability to complete the PID related tests. In other words, a large determinant of their ability to complete the PID tests might be their general lack of computer self-efficacy (CSE). Section 1 of the instrument therefore sought to benchmark participants' computer efficacy prior to the PID recognition tests and perception measurement, thereby enabling cross-disciplinary analyses of PID literacy by efficacy metric. The instrument made use of Howard's computer self-efficacy (CSE) measure (Howard, 2014). Howard's 12-item measure was determined to be preferable because it demonstrated improvements over popular alternatives, such Murphy et al. (Murphy et al., 1989), including a reduction in the number of questions and the incorporation of post-2010 computer language. Howard's measure also demonstrates improved internal reliability and psychometric properties (Howard, 2014, 2020). The 12-item measure includes 12 statements relating to computer efficacy, measured using a 7-point Likert scale (*Strongly disagree (1) — Strongly agree (7)*).

Section 2 was designed to elicit data about participants' recognition of PIDs and test the extent to which PIDs are understood, thereby contributing to our answering of RQ1 (*PID familiarity*) and RQ2 (*Identifying* *identifiers*). Four tasks each included screenshots taken from prominent scholarly publishers, articles, and repositories, each displaying several PIDs in context (Figure 3). Each of the four tasks challenged participants to identify the PIDs in the screenshots and indicate to which entity they pertained (e.g. *Publications on a publisher website or platform*, *Research data or open data*, *People*, etc.). Each challenge had only two correct responses. Participants were challenged on DOIs, ORCIDs, and Handle.net, with the PIDs contextualized differently in each task.

An additional six test challenges were included in this section. These each provided an example of a prominent PID types and requested participants to indicate the extent to which they were recognized and, if so, to which entity type they most associated them. For example, a DOI might be strongly recognized by the participant and most associated with *Publications on a publisher website or platform* and *Research data or open data*, and so forth. The creation of response options for the entities was informed by the 'landscape analysis' published in the literature (Cousijn et al. 2021). Participants were challenged on their recognition of the following PID types: DOI, ORCID, Handle.net, ROR, ISNI, and Uniform Resource Name (URN).

<center>*<Take in Figure 3>*</center>

Data to evaluate participants' perceptions of PIDs was captured and operationalized using the semantic differential measurement technique (Osgood, 1957; Snider & Osgood, 1969). The semantic differential measurement technique is one that has long been deployed to measure attitudes and perceptions towards concepts (or in Osgood's words, to 'measure meaning'). Each semantic differential scale consists of a series of bipolar adjective scales on which a participant responds, in relation to the object or concept being examined. Each adjective also corresponds to one of three 'semantic space dimensions': *evaluation, potency,* and *(oriented) activity*. These semantic space dimensions, or 'factors', are hypothesized to be used by humans in their assessment of almost all phenomena and, when combined with appropriate factor analysis, can yield a reliable measure of a participant's overall reaction to something (Stoutenborough, 2008). Though emerging from psychology, the semantic differential technique was quickly adopted for information science purposes (e.g. Allen and Matheson 1977; Geus et al. 1982; Katzer 1972) and continues to be deployed and refined to better understand a wide variety of information science research topics. Despite being a commonly used technique, it has been observed that few studies in information science use the approach optimally (Verhagen et al., 2015), with many erroneously conflating scale design and interpretation with Likert scales (Schrum et al., 2020).

In this instrument, PID perceptions were measured on the use of persistent identifiers against the following four concepts: *Scholarly Communications*; and the use of PIDs to refer to *People*, *Places*, and *Things* (Huber et al., 2016; Meadows et al., 2019). A total of 13 bipolar adjective pairs were created to assist in the measurement of PID perception against these concepts, using 9-points ranging from -4 to +4 to construct the semantic differential scale. These values were not displayed to participants. Each of the adjective pairs addressed the aforementioned 'semantic space' dimensions of Osgood's approach: *evaluation*, *potency*, and, (*oriented) activity* (Osgood, 1957). See Table 1. Osgood notes a degree of relative importance exists within the semantic space dimensions, with *evaluation* representing a more powerful dimension in human thinking than either *potency* or *activity* (Osgood, 1957). The conception and design of this section of the research instrument was therefore optimized to ensure both bipolar adjectives and semantic space dimensions were appropriately addressed. Data from this section contributes to our answering of our *PID perceptions* research question (RQ3).

*<Take in Table 1>*

Section 4 simply comprised five questions exploring participant's PID (re)use behaviour, as per the *Habits* research question (RQ4). This was a brief question section and attempted to elicit data on how PIDs were or were not being created, used, or reused by participants. The extent of PID creation among participants is a significant behaviour to measure; but reuse is arguably a more significant characteristic of PIDs (e.g. reusing PIDs that already exist within the PID graph).

## Data collection

Scholarly participants were recruited via social media (Twitter, Mastodon) and through established email lists. Email lists pertaining to open science, scholarly communications, and open repositories were targeted (e.g. code4lib, UKCORR). The memberships of these lists predominantly comprise individuals associated with the delivery of scholarly communications and research publishing support at scholarly organizations and who would have internal academic staff lists to whom the invitation to participate could be directed. The authors also circulated details of the participant invitation on local academic institutional lists. Only scholars who were outside the orbit of open science, scholarly communications and open repositories were invited to participate, since participants from these groups would inevitably demonstrate abnormal PID literacy relative to other scholarly groupings. Participation was restricted to only those scholars who were active in the publication or dissemination of research.

The data collection approach generated a convenience sample, a limitation to be discussed in a later section. Data collection occurred 26 May 2022—24 June 2022, with all data extracted shortly thereafter in .csv for cleaning, analysis, and visualization, all performed in standard spreadsheet software. A total of 106 academic users participated in the test, with the results of 27 removed during cleaning (i.e. invalid responses).

## Results

Tables 2 and 3 report basic demographic data on the nature of the test participants. Those included were drawn primarily from the Social Sciences (ss) (47%) and Physical Sciences (ps) (38%) but also, to a lesser extent, Life Sciences (ls) (15%). There were zero participants from the Health Sciences. Participants occupied the full spectrum of job role options available in the instrument, with the greatest proportion noted as being 'Professor / Reader', 'PhD Research Student', and 'Other'. The clustering of participants in 'Other' was unanticipated and suggests that the instrument lacked sufficient specificity in this area. Given that those participating were active in research publishing and dissemination, we could infer that some of those participants in this category occupied teaching fellow or associate positions, or were research / knowledge exchange associates, adjunct professors, and so forth.

Figure 4 summarizes the geographic origin of test participants, which were largely from the United Kingdom (57%), United States (16%), Germany and Canada (5%), and Italy (3%). A long tail of solitary participants came from countries including China, Brazil, South Africa, Norway, France, Austria, Czech Republic, The Bahamas, Croatia, Belgium, and Lithuania.

*<Take in Table 2 & Table 3>*

*<Take in Figure 4>*

### *Computer self-efficacy*

Recall that section 1 of the test instrument benchmarked CSE across participants using the specified CSE 12-item 7-point measure (Howard, 2014). Internal consistency of the scales was tested using Cronbach's alpha and demonstrated strong reliability ($\alpha = 0.94$). Data from the measure are summarized in Table 4. CSE results across the group revealed a moderate level of efficacy ($M = 4.98$; $Mdn = 5$) with a high level of variation around the mean ($SD = 1.52$). The highest possible total score for each participant in the CSE measure was 84. Measures of central tendency on this dimension ($M = 59.09$; $Mdn = 60$; $SD = 14.60$) highlights the large spread of reported computer efficacy across the participant group.

Table 4 also segments CSE data by the ASJC discipline grouping of participants. Life Sciences ($M$ = 4.84; *Mdn* = 5; $SD$ = 1.77), Physical Sciences ($M$ = 5.19; *Mdn* = 5.25; $SD$ = 1.53), and Social Sciences ($M$= 4.85; *Mdn* = 5; $SD$ = 1.41) all demonstrated considerable dispersion from the mean. Similarly, summarizing total CSE scores demonstrated dispersion but revealed that those in Physical Sciences (*ps*) demonstrated a slightly higher mean CSE score suggesting possible differences across discipline groupings ($M_{ps}$ = 62.07; $Mdn_{ps}$ = 61.50; $SD_{ps}$ = 15.19; $IQR_{ps}$ = 15.50; $M_{ls}$ = 57.57; $Mdn_{ls}$ = 63.50; $SD_{ls}$ = 17.09; $IQR_{ls}$ = 20.50; $M_{ss}$ = 58.24; $Mdn_{ss}$ = 60; $SD_{ss}$ = 12.59; $IQR_{ss}$ = 15.25). A one-way analysis of variance (ANOVA) ($\alpha$ = .05) of CSE scores across discipline groups was performed and post-hoc comparisons were performed using the Games-Howell post-hoc procedures. The Games-Howell test is noted by recent statistical research to be most suitable when data do not satisfy homogeneity of variances assumptions, with unequal sample sizes and unequal variances. (Rusticus & Lovato, 2014; Sauder & DeMars, 2019). The ANOVA reported no statistically significant differences.

Participants were further segmented by job role to observe possible CSE differences arising from academic duties, experience, etc. Summary data are set out in Table 5. Here we can observe that participants belonging to specific job roles reported greater efficacy than others. Interestingly, Lecturers ($M$ = 47.67; $SD$ = 19.09; $IQR$ = 27.25) and Professors / Readers ($M$ = 54.30; $SD$ = 13.28; $IQR$ = 16.75) reported the lowest mean levels of efficacy but with greater levels of variation, while Research Assistants ($M$ = 75; $SD$ = 9.85; $IQR$ = 9.5) and Postdocs ($M$ = 67.80; $SD$ = 5.81; $IQR$ = 4) reporting the inverse. Despite observable variances between groups, a one-way ANOVA ($\alpha$ = .05) of CSE scores across job role groups reported no statistically significant differences.

### PID recognition challenges

Data arising from the first batch PID recognition challenges are set out in Table 6. These data relate to the four tasks within section 2 of the instrument, which challenged participants to identify the PIDs in screenshots and indicate which entities they identified. Each of the four tasks had only two correct responses. The correct responses in each task are indicated in Table 6 by an asterisk; those without an asterisk denote erroneous responses to the task.

A significant proportion of participants were successful in the challenges and correctly interpreted PIDs, with a high and low rate of 86% and 61% for the two components of each specific challenge respectively. However, we can note that some participants failed the task #1 challenge, with ~ 15% of participants ($n$ = 12) failing to correctly interpret the PIDs. The failure rate was higher for task #2, with ~ 22% ($n$ =17) and ~ 34% ($n$ = 27) observed. Rates for tasks #3 and #4 noted similar failure rates, with ~ 24% ($n$ = 19) and ~ 25% ($n$ = 20) for task #3 and ~ 14% ($n$ = 11) and ~ 39% ($n$ = 31) for task #4.

These results can be segmented by discipline (Table 7) and job role (Table 8), within which we can observe the overall performance of participants in these challenges. Participants from Social Sciences performed least successfully, achieving a score of ~ 66%, ergo 34% of responses were incorrect. The maximum score that any single participant could achieve across the four challenges was 8; yet those in Social Sciences demonstrated huge levels of dispersion ($M$ = 5.24; $SD$ = 2.89; $IQR$ = 5), exposing a wide variation in the success of Social Science participants in the task. This contrasts with Life Sciences participants, whose overall performance yielded a higher score (~ 85%), with individual participants demonstrating greater homogeneity ($M$ = 6.83; $SD$ = 1.40; $IQR$ = 1.25). Those from Physical Sciences were found to sit between Life Sciences and Social Sciences, with an overall success score of 70%. The differences appeared notable but were nevertheless tested. Owing to the statistical nature of the data group, a Levene test was performed and found that the homogeneity of variance assumption was not satisfied ($p$ = > .001). A one-way ANOVA (with Welch F test - α = .05) of participants' scores by discipline was therefore performed, indicating a statistically significant difference ($F$(2, 39.52) = 3.86, $p$ < .03) between the performances of disciplines groups. Post-hoc comparisons using the Games-Howell procedure (described previously) confirmed a significant difference between the performance of Life Science and Social Science participants only ($p$ < .04).

*<Take in Table 7 & Table 8>*

Summary results by job role are set out in Table 8, alongside summary results by discipline and measures of central tendency provided for individual performance on recognition tests by discipline. Participants identifying as Research Assistant demonstrated the highest mean score in the PID recognition challenges (~ 96%) while those identifying as Research Fellow scored lowest (50%). A huge range in individual scores can be observed from some SD figures. For example, some of those identifying as Research Support ($M$ = 64.06; $SD$ = 3.31; $IQR$ = 5.50) got none of the challenges correct. Following verification of data as demonstrating significant variance (Levene, $p$ = > .02), a one-way ANOVA (with Welch F test - α = .05) of participants' scores by job role suggested a significant difference ($F$(7, 19.65) = 3.11, $p$ < .03); but post-hoc comparisons using the Games-Howell procedure failed to confirm this ($p$ > .05). Despite many participants identifying the PIDs correctly in the challenges, a wide variety of additional erroneous responses were often provided (Table 6). Using the same statistical procedures as previously, a statistically significant difference was found between job roles when erroneous responses were provided ($F(7, 21.53) = 5.98, p < .001$), in which those designated as PhD Students ($p < .02$) and Other ($p < .04$) were more likely to offer erroneous responses, and in greater number.

Recall that an additional six test challenges were included in this section of the research instrument. These each provided an example of a prominent PID type and requested participants to indicate the extent to which they were recognized and, if so, to which entity type they most associated them. It was anticipated that some of these types would be better known to participants than others. Data are summarized in Figure 5 and Table 9.

The proportion of participants strongly recognizing PIDs was unsurprisingly highest for DOIs and ORCIDs, at 68.35% and 63.29% respectively, and lowest for URNs (5.06%) and ISNIs (5.06%). Test challenges relating to other PID types generated mixed responses. For example, recognition of Handle as a type of PID generated considerable spread across the response options (*Do not recognize* = 34.18%; *Unsure* = 12.66%; *Somewhat recognize* = 16.46%; *Strongly recognize* = 22.78%). In indicating to which entity type participants most associated the PIDs, those who recognized the PIDs provided a spread of responses, summarized in Table 9. Owing to the nonparametric nature of data in this section, Kruskal-Wallis tests were performed between discipline and job role groupings. Statistically significant differences were observed by discipline in the recognition of Handles ($H(2) = 8.14, p < .02$) and URNs ($H(2) = 6.08, p < .05$) only, to varying levels of significance. Differences were also confirmed in the recognition of ORCID ($H(2) = 12.77, p < .02$), ROR ($H(2) = 13.96, p < .03$) and URNs ($H(2) = 12.11, p < .03$) by job role.

*<Take in Figure 5 & Table 9>*

### *PID perception measurement*

Combined with the PID recognition challenges, the PID perception measurement section represented the next significant portion of the research instrument. Table 10 presents the 'factor scores' for the ratings of PID concepts across all participants. Factor scores are derived from averaging the results for each bipolar adjective pair by the number of semantic dimension subjects present. For example, *potency* features four times as a semantic dimension (see Table 1) and therefore '4' is the denominator.

From Table 10 we can report that participants indicated their lowest collective response to the *activity* dimension, across all PID concepts ($M = 0.40$). Within this dimension, the use of PIDs to identity *People* — an allusion to PID types such as ORCID — was considered most positively ($M = 0.72$) albeit low in comparison to the scores attained in other dimensions. Conversely, use of PIDs within *Scholarly Communications* ($M= 0.55$) or to identify *Places* (*M* = 0.21) or *Things* (*M* = 0.13) was perceived more negatively in the *activity* dimension. The most positive perceptions were observed in the *potency* dimension, yielding a factor mean of 1.70, in which the scores for the tested concepts (excepting *Places*) were highest. Here the perception of PIDs in *Scholarly Communications* (*M* = 2.09), *People* (*M* = 1.87), and *Things* (*M* = 1.81) were perceived most positively and higher than the *evaluation* dimension in most cases.

<Take in Table 10 & Table 11>

Results for PID perception measures across discipline groupings by semantic dimension are provided in Table 11 and reveal some disciplinary differences. Participants from Life Sciences were generally found to perceive the use of PIDs more favourably, particularly in relation to the *People* concept, where participant ratings across all dimensions was > 3, considerably higher than other disciplines. The exception to this positive appraisal from Life Sciences was in relation to the *Place* concept, where some of the most negative perceptions across all concepts, dimensions, and disciplines was observed. From across all three disciplines, those from Physical Sciences displayed a generally more negative perception towards PIDs, with generally lower factor scores across all the tested concepts and their semantic dimensions. For example, the combined concept score (CCS) across all concepts — the overall perception measure — was lower than either Social Sciences or Life Sciences participants and was similarly lower based by dimension, with factor means of 1.27, 1.24 and 0.22 for *evaluation*, *potency*, and *activity* dimensions respectively.

Osgood and others describe the how semantic distance can be charted using the distance notion (Osgood, 1957; Rosnow, 2000; Snider & Osgood, 1969; Stoutenborough, 2008). This can assist in identifying specific semantic dimensions or bi-polar adjectives within the semantic scales which have triggered specific responses, and how those responses may relate to others. Figure 6 provides four semantic distance charts, one charting data across all participants (A) and three by discipline (B, C, D). Data supplementing these charts are provided in Table 12. The semantic distance between concepts (*D* = *distance*) — derived using the generalized distance formula — is provided in Table 13.

<Take in Figure 6>

<Take in Table 12 & Table 13>

The significance of the chart profiles will be explored in more detail within the discussion section; suffice to state that visual inspection of the Life Sciences chart (D) demonstrates a greater propensity for extreme perception differences across semantic dimensions as well as on the tested PID concepts. By contrast Social Sciences participants demonstrated generally consistent perceptions across the concepts, with fewer extremities noted. A tendency for the perception of different PID concepts to track each other can also be observed in Social Sciences. Within Physical Sciences, we note that *People* and *Places* track each other closely while *Scholarly Communications* and *Things* deviate from this chart profile, displaying comparatively irregular perceptions.

A Wilcoxon signed-rank test can be performed to examine *D*, as detailed in Table 13. This was used ($\alpha$ = .05) to determine whether *D* between specific discipline groupings were significant. Results are summarized in Table 14. Significant *D* differences were observed between *Scholarly Communications* and *Things*; Physical Sciences and Social Sciences demonstrated a significant distance to Life Sciences, with the former reporting significance at $\alpha$ = .05 ($T$ = 3.58, $z$ = 2.11, $p$ = 0.02) and the latter at $\alpha$ = .01 ($T$ = 3.74, $z$ = 2.19, $p$ = 0.01). Significant results were also observed for *D* between *Places* and *Things*, with the discipline groupings of Physical Sciences and Social Sciences notable ($T$ = 2.52, $z$ = 2.56, $p$ = 0.01).

*<Take in Table 14>*

### PID (re)use habits

Recall that our research instrument concluded with five simple questions eliciting participant's PID (re)use behaviour. On their (un)familiarity with using PIDs in scholarly communications, as measured on a 9-point bi-polar adjective scale (*Unfamiliar — Familiar*), participants reported themselves to be generally familiar, with median scores inferring considerable confidence ($M$ = 7.68; $Mdn$ = 9.00; $SD$ = 1.81) (Table 15). On their understanding of the purpose of PIDs (*Unknowledgeable — Knowledgeable*), participants also considered themselves generally knowledgeable ($M$ = 7.29; $Mdn$ = 8.00; $SD$ = 1.89), although with less confidence than their familiarity. This observation held across all discipline groupings and all job groups, except 'Others' which reported equal familiarity and knowledge (Table 16). Results indicate less confidence from Physical Sciences relative to Social Sciences and Life Sciences (Table 15) but also within specific job groups, most notably Research Assistant (Table 16). Results indicate that Social Sciences participants considered themselves most familiar and knowledgeable about PIDs. Incrementally higher levels of dispersion from the mean can also be observed as job 'seniority' declines. For example, (un)familiar and (un)knowledgeable for Professor / Reader ($SD$ = 0.41; 0.41) and Other ($SD$ = 3.25; 3.25).

*<Take in Table 15 & Table 16>*

Participants' views on the purpose(s) of PIDs are detailed in Table 17, with > 80% of participants noting that they exist to ensure the persistent and unambiguous citation of scholarly entities on the web. The use of PIDs as a way of mitigating 'link rot' or 'reference rot' in scholarly communications was also noted (> 78%). The importance of PIDs in contributing to global scholarly graphs was, however, noted as a consideration for only 49% of participants.

Specific questions on the creation and (re)use of PIDS over the past 4 years indicated that close to 74% of participants reported creating a PID to identify a preprint or accepted author manuscript, with 54% reporting that it had been subsequently reused (Figure 7). Many participants indicated their creation and reuse of PIDs for people, at 57% and 49% respectively. The creation and reuse of PIDs for identifying (open)research data was also noted, at 36% and 35% respectively, with a longer tail of responses relating to PIDs for software, projects, and so forth. Few participants reported use of PIDs to identify research instruments (3%) or research equipment (1%).

*<Take in Table 17 & Figure 7>*

## Discussion

Several interesting discussion points emerge from the findings. We use the previously introduced research question labels to structure the discussion: *PID familiarity* (RQ1), *Identifying identifiers* (RQ2*), PID perceptions* (RQ3), and *Habits* (RQ4).

### *PID familiarity & 'identifying identifiers'*

Firstly, though results from the PID recognition challenges were varied, it is possible to conclude that many participants across all discipline groups failed to satisfy the components of what we defined as PID literacy. The first four PID recognition tests challenged users to identify common PID types within context (e.g. within real-world published articles). A generous interpretation of data indicates that even in Life Sciences, where participants' performance was best, circa 15% of responses were still incorrect. Rates were much lower in Physical Sciences (30%) and Social Sciences (34%), highlighting that almost one third of participants in these discipline groupings were unable to correctly identify PIDs, as they might commonly be presented within a scholarly journal article or repository. In the case of Social Sciences, we can also conclude that this result is statistically significant insofar as it stresses a distinct disciplinary divide between the aptitude of participants from Life Sciences and Social Sciences in this respect.

Closer examination of the results helps us to observe a level of uncertainty among scholarly participants when challenged in the four tasks. Despite many participants identifying the PIDs correctly in the challenges, a wide variety of accompanying but erroneous responses were also often provided. For example, data for task #1 — the correct responses for which were 'Publications (on a publisher website or platform)' (i.e. DOI) and 'People (e.g. authors, editors, PIs, etc.) (ORCID) — revealed that some of those who identified the two correct responses also included additional incorrect responses (equivalent to 21.6% of cases). It is worth adding that the responses to the challenge in task #1 included two of the most widely used PIDs (i.e. a DOI identifying a publication and ORCIDs identifying authors of that publication). That some participants considered these PIDS — that were contextualized within a real-world academic article — to also identify projects, research grants, a publication on a repository, etc. is a concerning indicator of PID literacy. Suffice to state that many participants who submitted correct responses were similarly uncertain about whether these PIDs identified other entities too. This is significant because such responses inadvertently reveal that these individuals did not appear to understand the notion of PIDs as *unique* identifiers.

One possible explanation is that social scientists typically publish less often (Hicks, 2013) and are more likely to view long-form publications as a vehicle for research dissemination (De Filippo & Sanz-Casado, 2018), resulting in reduced exposure to the emerging centrality of PIDs to scholarly communications. They are also less likely to be in receipt of research funding and are less likely to generate research data (Curty, 2016; Jarolimkova & Drobikova, 2019), both of which might ordinarily bring them into contact with adjacent PID concepts; for example, FAIR data (David et al., 2020), research funders (Lammey, 2020), software (Li et al., 2016), research instruments (Stocker et al., 2020), and so forth. Such an explanation may prove unsafe in the long-term as recent evidence suggests that the publication behaviours of social scientists may be evolving in line with a global growth in national research evaluation exercises (Savage & Olejniczak, 2022). However, this does not explain the performance of participants from Physical Sciences. Though their performance was not found to be statistically significantly different to Life Sciences, visual inspection of data indicates that they were only marginally better than those in Social Sciences.

While disciplinary differences were clearly observable, disentangling job role effects from these is difficult without a larger sample to better represent individual job roles. No statistically significant differences between job role performance were detected. We can at least infer, perhaps unsurprisingly, that users' computer self-efficacy (CSE) appears to have little connection to PID literacy. The CSE benchmarking and the analysis of the results found no significant differences between disciplinary or job role groups, suggesting that participants represented a relatively consistent level of computer efficacy. Their level of PID literacy, as measured in this research, therefore appears independent of this efficacy. However, we can note in the analysis of erroneous responses (as segmented by job role) that PhD Students were particularly more likely to offer erroneous responses and that the number of these erroneous responses tended to be higher. The embryonic nature of a PhD student's research career dictates a relative lack of experience with scholarly publishing and scholarly infrastructure, which may explain what was observed (Hatch & Skipper, 2016). It may also be reflective of the levels of digital illiteracy observed in the reviewed literature. The recent emergence of 'researcher development' initiatives at research organizations has highlighted the need to better equip postgraduate researchers for their research career (Rospigliosi & Bourner, 2019), with useful examples emerging from research library contexts where training has coalesced with aspects of information literacy education (Fazal & Chakravarty, 2021). Though such initiatives are likely to assist researchers in navigating many aspects of the publication lifecycle and stress digital scholarship competencies, it is conceivable that the emerging centrality of PIDs to scholarly communications has been absent from 2020s teaching content.

Results from the additional six test challenges appeared to confirm our findings from the previous four challenges. Low recognition of esoteric PID types was expected (e.g. URNs and ISNIs). While DOIs and ORCIDs were the most strongly recognized PID types, it is somewhat surprising that PID types of such ubiquity and visual distinctiveness were recognized by only 68% and 63% of participants. This may signify that some researchers display a disconnection with two of the longest standing and most widely used PID types, perhaps because of their publication culture (e.g. publication venues do not support DOIs or ORCID) and, as statistically confirmed, their job role precludes them from experiencing them (e.g. their role is such that they lack exposure to these types). This, however, is an unsatisfactory explanation. At time of writing ORCID penetration within research organizations is high. More than 10 million ORCIDs have been added to the ORCID registry at time of writing (Petro, 2020). Research funders and national research assessment exercises increasingly consider ORCIDs to be mandatory for research active staff (Choraś & Jaroszewska-Choraś, 2020). Most participants would be expected to therefore be ORCID literate, have an ORCID, or at least have recognition of them.

Uncertainty continues within this context too. For example, in indicating to which entity types the PIDs were most associated, DOIs — which can theoretically be coined for any web entity, but which tend to have specific applications within scholarship — were most noted as being associated with publications (77%), publications on a repository (37%), research data (19%), software (3%), and so forth. But ORCIDs were most noted as identifying people (68%), but also publications (13%), publications on a repository (14%), research data (5%), and other entities. This would indicate that even those individuals who recognize ORCIDs and perhaps even have an ORCID, are unsure of their ultimate purpose. As we shall see in the following section, PIDs as a mechanism to identify 'people' were nevertheless perceived positively. Individuals therefore perceive ORCIDs to be a positive thing, irrespective of any confusion surrounding how they might work or cultural disciplinary differences.

The low recognition of Handles is notable, especially being statistically linked by discipline grouping (alongside URNs). Like DOIs, Handles (Handle.net) can be coined for virtually any web entity and forms the basis of a number of identifier types, such as RAiD (Janke et al., 2017). They are perhaps most commonly used in open repository infrastructure to ensure persistent identification of open research content (e.g. manuscripts of research articles, research data, etc). The huge volume of open research content now being served by repositories (de Castro & CESAER, 2022) and their centrality to the wider open research ecosystem, would suggest that users' familiarity with Handle should be greater. Interpreting this lack of familiarity may be disciplinary, as statistically inferred. But it may also infer that some participants would tend not to use Handles when citing such content, instead relying on more transient URLs, and have therefore tended to ignore Handles when they are provided by repositories. This might be because they have not understood their purpose, although may also reflect the findings of the reviewed literature: that some scholars' struggle to correctly cite digital entities. The increased preference for DOIs to be used by repositories instead (cOAlition S, 2022) is in part because they are considered more recognizable to users than Handles and therefore better contribute towards initiatives such FAIR data (Dunning et al., 2017) and Pubfair (Ross-Hellauer et al., 2019). Their data contribution to the PID graph is also currently superior (Cope, 2021).

## *PID perceptions*

Measuring participants' perceptions of PIDs is important to our research motivation since such perceptions are likely to influence future PID user behaviour. Despite the findings associated with *PID familiarity* (RQ1) and *Identifying identifiers* (RQ2), it is significant to observe that participants perceived the use of PIDs to be generally a positive thing. This is especially clear in relation to the semantic dimensions of *evaluation* and *potency* across all the tested concepts, with *Scholarly communications* and *People* enjoying the highest perception ratings. Of the concepts tested, these are likely the most familiar to participants and therefore enjoy a level of demonstrable utility that *Places* and *Things* do not, owing to their association with the identification of scholarly publications and authors. However, it demonstrates that PID perceptions, in terms of participants' evaluative attitudes and the potency with which these attitudes are held, is generally high. This observation is less true when we consider the *activity* semantic dimension, where factor scores were much lower. The *activity* dimension is an indicator that while positive perceptions exist on evaluative and potency terms, PIDs are perceived less favourably when action is required. For example, they are closer to being perceived by participants as 'laborious' and 'complex'. The significance of negative perceptions around *activity* become more obvious at a disciplinary level, particularly in Physical Sciences and Life Sciences, which held some of the most negative perceptions on this semantic dimension.

Of the PID concepts measured for perception *Places* was universally perceived most negatively, irrespective of discipline. The persistent identification of places represents a typical PID application; for example, to identify research organizations such as universities or research funders. Interpreting this finding is difficult without additional qualitative data but we may speculate that the notion of PIDs for places was considered too abstract for some participants, whereas PIDs for *People* or *Things* was considered more tangible and more relatable to scholarly practice.

Results of PID perception measurements across disciplines revealed compelling differences. Overall, Life Sciences participants demonstrated the most favourable perceptions of PIDs, perceiving the use of PIDs to identify *People*, *Things* and within *Scholarly communications* most positively; but they also demonstrated a greater inclination towards extreme perceptions on specific semantic dimensions or concepts. This was clearly observable from the corresponding semantic distance chart (D). For example, the *People* and *Places* concepts yielded a perception rating of 3.75 and -1.00 respectively for the bi-polar pair, '*Unintuitive—Intuitive*' (belonging to the *evaluation* dimension). More extreme differences were observed across bi-polar scales belonging to the *activity* dimension (e.g. *Complex—Simple*, *Laborious—Effortless*, *Difficult—Easy*). The concepts of *People* and *Places* too were, in general perception terms, far wider apart than either Physical Sciences, Social Sciences, or all participants taken collectively. Also notable is the extent to which the profile of the semantic distance chart for *People* and *Places* covary, with the perception of one tracking the perception of the other but with large distances between them. In other words, participants in Life Sciences were the most likely to perceive PIDs for *People* positively but simultaneously the most likely to perceive PIDs for *Places* negatively. Conclusions about Life Sciences were corroborated with the measurement of *D*, with *D* calculated to be considerably higher across all but one concept when compared to other disciplines. That *D* was found to be statistically different to Physical Sciences and Social Sciences for specific concepts is noteworthy and highlights a potential perceptual distinction between the disciplines.

An explanation for this finding is difficult without additional qualitative data from Life Sciences participants; but we may suggest that it is fuelled by the rapid prominence of PIDs within Life Sciences. Whether through the recent but rapid growth in pre-printing (Johansson et al., 2018; Sarabipour et al., 2019), a phenomenon accelerated by the COVID-19 pandemic (Fraser et al., 2021; Majumder & Mandl, 2020) and joined by an accompanying consciousness of FAIR data sharing approaches (Austin et al., 2021). Or the recent proliferation of PID types for entities such as samples (Bandrowski & Martone, 2016; Lehnert & Klump, 2018), specimens (Hardisty et al., 2021), equipment (Haak et al., 2018), materials (Lehnert et al., 2019), instruments (Plomp, 2020), and so on; it is conceivable that the increased visibility of PIDs has exposed these researchers to their function. The increased exposure may also help to explain why stronger perceptions were found on specific concepts as well as specific semantic dimensions (i.e. *activity*).

By contrast Social Sciences participants demonstrated generally consistent perceptions across the tested concepts, with fewer extremities noted. A cursory visual inspection of the corresponding semantic distance chart (C) shows this consistency, relative to both Life Sciences and Physical Sciences; confirmed in most of the calculations of *D*. But we should also acknowledge that *activity* was again the dimension attracting least positive perceptions. The differences of *D* between Physical Sciences and Social Sciences for *Places* and *Things*, though statistically significant, appear to be significant because of the generally lower perception ratings displayed by Physical Sciences to PIDs. Excepting an isolated outlier for *Places* on the *activity* dimension, those from Physical Sciences displayed a generally more negative perception towards PIDs across all the tested concepts and their semantic dimensions. Such a finding is counterintuitive since it might be expected that Physical Sciences participants would be more favourably disposed to PIDs than those in Social Sciences, particularly as the latter group demonstrated lower familiarity with PIDs in the recognition tests.

## *Habits*

In interpreting results on PID (re)use habits we can immediately observe that participants indicated high levels of familiarity and knowledge with PIDs (optimum levels if only median values are considered) but that this confidence was in no way reflected in the results of their corresponding PID recognition challenges, described in *PID familiarity* and *Identifying the identifiers* sections above. In other words, participants considered themselves to be generally familiar and knowledgeable about PIDs, despite simultaneously demonstrating that they were often both unfamiliar and unknowledgeable about them in practice. This is especially true of Social Science participants. But on these data, it appears that interpretation of the results by job role may be more meaningful, since far greater variation between roles than between disciplines is observable. Notwithstanding the reported high levels of familiarity and knowledge as per mean and median values for particular job roles (e.g. Postdoc, Research Support, Other), the high dispersion from the mean suggests high data spread — spread which appears to grow as job seniority declines. The incrementally higher standard deviation as job roles become less senior is again likely indicative of uncertainty among some participants. We should, however, acknowledge that it also suggests those at Professor / Reader or Lecturer level considered themselves to be more familiar and knowledgeable, and to be more consistent in this perception than, say, a participant from Research Support.

The more PIDs are reused, the more that new nodes can be defined by existing PIDs, thereby enriching the graph. Our definition of PID literacy emphasizes the importance of users' understanding of when PIDs should be used and when they should be reused. PID reuse is critical to the vitality of the resulting scholarly graph (Dappert et al., 2017). Typical enrichment might take place surrounding, say, an individual (via a person PID such as ORCID) or a research dataset (via explicit dataset linking from associated publications using a DOI). The recent rise in preprint publication (Lin et al., 2020) presents a common point of PID creation for many researchers, since preprint deposit will often entail PID minting (i.e. DOI). It was therefore unsurprising to find that this was the most highly reported context for PID creation, even if reuse was some 20% lower. The creation of PIDs for research data may appear low at 36% but demonstrated high levels of subsequent reuse, most probably because research data are likely to form the basis of more than one associated research publication and/or be used repeatedly within a wider research project (Borgman & Wofford, 2019).

Accepting that only a significant minority of the study participants would have the need to create a PID for research data, software or instrument, the proportion of participants who indicated reusing a PID for people was disappointing. Though several PID types exist for defining people, our research instrument was clearly alluding to ORCID, with which most participants would be to be familiar. That fewer than half of participants indicated they had reused a PID for people was therefore a noteworthy finding, and one that should be the subject of further research. It is possible that the description of 'PIDs for people' within the research instrument was too opaque for some participants to make the connection with ORCID, resulting in under reporting. However, it could also be a legitimate finding influenced by the disciplinary context of some participants. In the case of preprint PID reuse, is it that these PIDs became superseded by an accepted publication, rendering the PID defunct for reuse? Understanding participants' PID reuse behaviour is something which requires an additional, separate piece of qualitative research, preferably through user interviews or protocol analysis approaches.

It is possible that these lower levels of creation and reuse are linked to responses surrounding the purpose of PIDs. PIDs perform a wide number of functions but no specific 'PID purpose' was considered universally relevant by participants (Table 17), thereby demonstrating a degree of PID illiteracy among participants. The notion of PIDs as enabling 'persistent and unambiguous citation of scholarly objects' was a purpose that only circa 80% considered core. The importance of PIDs in contributing to global scholarly graphs was therefore considered even less relevant, at 49% of participants. With no widely accepted understanding of why PIDs exist and why they are necessary, it seems apparent that unsatisfactory and erratic (re)use of PIDs will always occur among scholarly users.

## Future research and limitations

Our research suffers from several limitations, which also stimulate further research. Firstly, not all possible variables were tested or discussed in this work, due to the limits of space. Secondly and more importantly, the exploratory nature of the research topic, and the consequent research instrument designed for this study, could not elicit all possible data needed to better understand the research area. Operating as a remote instrument, the instrument design, such as the PID challenges, were necessarily artificial and could not replicate the control of laboratory conditions. Nor could it assist us in better understanding why participants performed the way they did. The instrument was therefore satisfactory at surfacing perception data, as well as identifying differences between groups, but less satisfactory at understanding why these perceptions or differences existed. There is consequently a need for further research to address this weakness, ideally incorporating mixed methods. Such further work might use a smaller cohort of participants by studying users within a controlled task-based setting perhaps using, for example, protocol analysis (or 'think aloud') (Ericsson & Simon, 1993) or stimulated recall user study approaches (Lazar et al., 2014), thereby generating rich qualitative data which could be mined for insight. The resulting qualitative data would likely enable improved understanding of user uncertainty surrounding PIDs (especially ORCIDs) and shed light on why users' level of PID literacy and perception can vary across disciplines and job roles. This would especially assist in better differentiating between levels of PID literacy and PID perceptions at a disciplinary and job level, and under which circumstances each becomes a relevant factor in scholars' PID literacy. It may also provide a sound basis for proposing a model of PID literacy, capable of specifying the suite of competencies today's digital scholars require in order to interact meaningfully with PIDs.

The disciplinary differences detected between participants also prompts greater benchmarking of PID literacy across disciplines, perhaps generating a maturity model or PID literacy model to guide the PID expectations emerging from research funders and proponents of open scholarly infrastructure. Of course, additional work would be welcomed to corroborate statistically significant findings, as well as consolidate findings on users' PID perceptions.

Finally, the recruitment and sampling of participants was designed to accommodate the exploratory, unfunded nature of the work. This approach may, however, have introduced a level of data bias, potentially limiting the generalizability of the results (Bornstein et al., 2013; Rusticus & Lovato, 2014). It can certainly be noted that participants from Health sciences were not included and that unequal group sizes were also used, although steps were taken in subsequent data analysis to control for this. We nevertheless remind the reader of the small sample size used and that the results of statistical tests should be accepted with caution. Suffice to state, it would be useful for future research to combine the need for additional qualitative evidence with the equal recruitment of participants from all disciplinary areas.

## Conclusions

The increased use of PIDs in open scholarly infrastructures is evident to those who interact with it. Trends indicate that a growing number of PID types will soon emerge to better enable the persistence, discovery, citability, traceability, and verification of scholarly research entities (Hardisty et al., 2021). What our exploratory research shows is that many scholars, even allowing for disciplinary or job role differences, may demonstrate inadequate levels of PID literacy. They are either uncertain about the function of PIDs and what they might identify or cannot discriminate between different PID types, even when they are contextualized within real-world examples. While this study exposed participants to lesser known PID types, such uncertainty and confusion was found to exist in relation to types that would otherwise be considered dominant or typical, such as ORCIDs, or DOIs for academic objects such as journal articles or research datasets. Indeed, despite self-reporting high levels of PID cognizance, irregular patterns of PID literacy and certainty were found to exist across all participants, though statistically significant disciplinary differences were observed in some instances, notably between Life Sciences and Social Sciences. This work therefore contributes to our understanding of scholars' PID literacy and functions as an alert to those pioneering PID-centric scholarly infrastructures, that a significant need for training and outreach to active researchers remains.

It may be recalled that our research was motivated from a data perspective. PID-centric scholarly infrastructures are largely predicated on the notion of the PID graph. But without addressing scholars' levels of PID literacy it is probable that, at least initially, the resulting graph will not only lack relational depth and ergo inferential potential but suffer empty nodes owing to low PID (re)use by researchers. The educational and training offered by learned societies, research funders, open research and scholarly communications teams based at academic higher education organizations should therefore also be informed by this work.

Despite the low levels of PID literacy found by this study, our work nevertheless found scholars' perceptions of PIDs to be generally positive. By applying semantic differential techniques, we discovered that positive perceptions of PIDs within scholarly ecosystems were offset by some pronounced disciplinary differences, as well as higher levels of aversion to PIDs in specific use cases. Negative perceptions were found to exist when concepts associated with PIDs were measured on an *activity* semantic dimension. These perceptual insights should inform future technical approaches to the implementation of PIDs; that, scholars perceive PIDs positively in evaluative and potency terms, as a mechanism to support their work, but view the actions or activities involved in PID (re)use or creation less favourably. While this exploratory work was motivated from a data perspective and its implications for the PID graph, it also offers a valuable snapshot of academic users' digital scholarship competencies and therefore contributes to the wider literature on information literacy. It also provides useful perceptual insights into academic thinking around the increased use of PIDs in scholarly ecosystems.

## Data availability statement

All data and research instruments underpinning the research documented in this article are available from:

https://doi.org/10.17868/strath.00083073

## Acknowledgements

This study was approved by the Department of Computer & Information Sciences Ethics Committee, University of Strathclyde (ID: 1804).

274

275

276
277
278

279

280

281

282

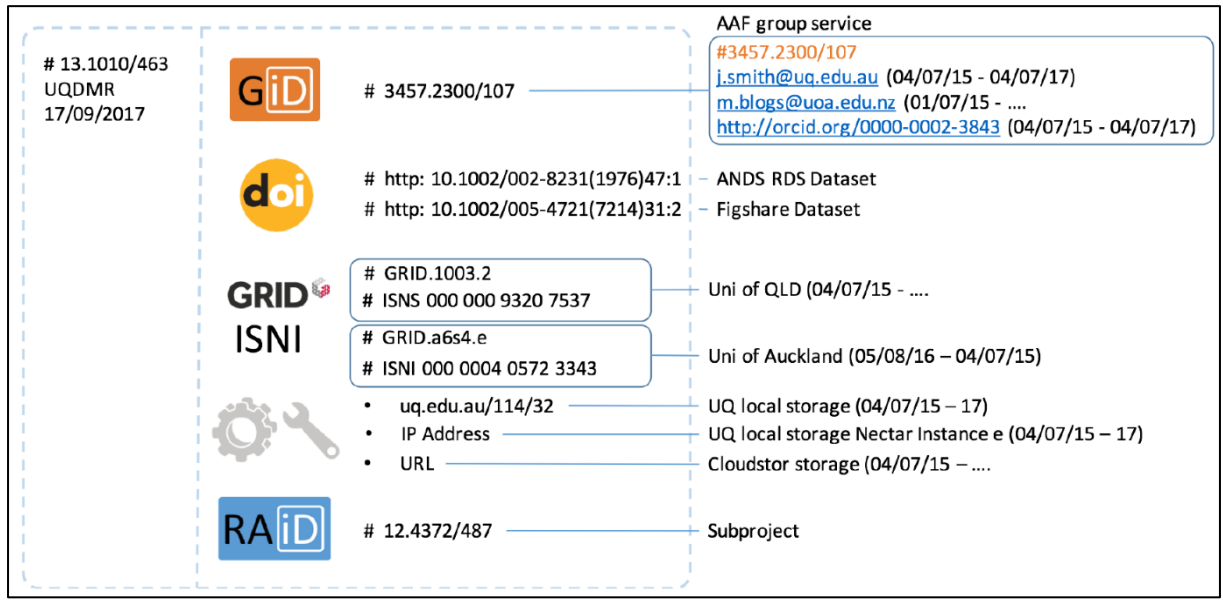

283

284 *Figure 2: Diagrammatic of a RAiD a 'metadata envelope for an example research activity (i.e. project), containing*
285 *associative relations via PID to related research entities. Diagram from Janke et al., 2017, used under Creative Commons*
286 *Attribution International License 4.0 (CC-BY).*

287

288

Above is a screen snippet of an article published by a well-known journal title.

After examining the screen snippet, please indicate below to which (if any) the highlighted persistent identifiers refer. (tick as many or few as you think is necessary)

Publications (on publisher website or platform)

Publications (on repository)

Research data or open data

Research grants

*Figure 3: Example of PID recognition test questions as used within Section 2 of the research instrument.*

293

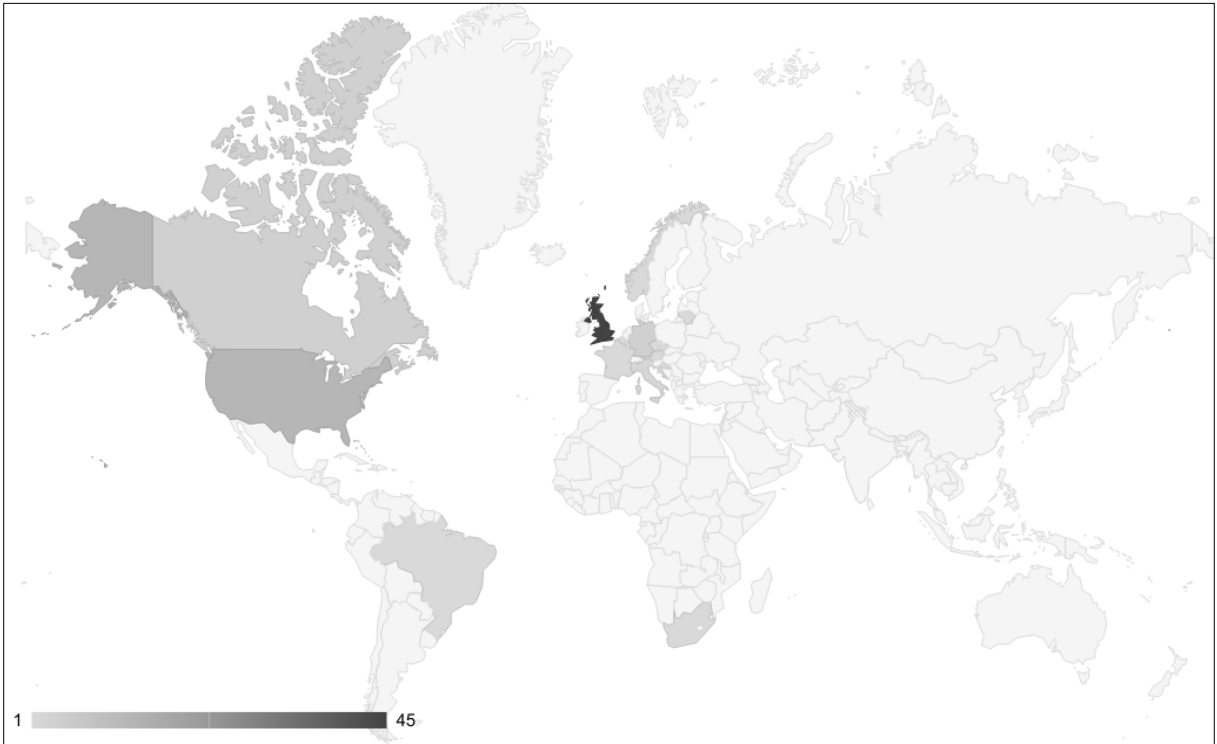

294
295

*Figure 4: Geographic origin of test participants visualized as a map summary.*

296

297
298    *Table 1: Summary of adjective pairs used in Section 3 of the research instrument, and their mapping to Osgood's semantic*
299    *space dimensions.*

| Semantic space dimension | Adjective pair |
|---|---|
| *Evaluation* | Bad - Good |
| *Potency* | Unimportant - Important |
| *Activity* | Complex - Simple |
| *Evaluation* | Unintuitive - Intuitive |
| *Evaluation* | Foolish - Wise |
| *Potency* | Unscientific - Scientific |
| *Activity* | Laborious - Effortless |
| *Evaluation* | Useless - Valuable |
| *Evaluation* | Unintelligible - Intelligible |
| *Potency* | Abstract - Concrete |
| *Activity* | Difficult - Easy |
| *Evaluation* | Negative - Positive |
| *Potency* | Unnecessary - Necessary |

300

301

302

303     *Table 2: Academic test participants categorized by discipline, using the All Science Journal Classification (ASJC).*

| Origin discipline of academic participant (by ASJC) | *n* | % |
|---|---|---|
| Physical Sciences (includes: Chemical Engineering, Chemistry, Computer Science, Earth and Planetary Sciences, Energy Engineering, Environmental Science, Material Science, Mathematics, Physics and Astronomy, Multidisciplinary) | 30 | 38 |
| Health Sciences (includes: Medicine, Nursing, Veterinary, Dentistry, Health Professions, Multidisciplinary) | 0 | 0 |
| Social Sciences (includes: Arts and Humanities, Business, Management and Accounting, Decision Sciences, Economics, Econometrics and Finance, Psychology, Social Sciences, Multidisciplinary) | 37 | 47 |
| Life Sciences (includes: Agricultural and Biological Sciences, Biochemistry, Genetics and Molecular Biology, Immunology and Microbiology, Neuroscience, Pharmacology, Toxicology and Pharmaceutics, Multidisciplinary) | 12 | 15 |
| **Totals** | **79** | **100** |

304

305

306

307

*Table 3: Test participants as categorized by academic job role.*

| Academic participant job role | *n* | % |
|---|---|---|
| Professor / Reader | 10 | 13 |
| Lecturer | 6 | 6 |
| Research Fellow | 7 | 9 |
| Research Assistant | 3 | 4 |
| Postdoc | 5 | 7 |
| PhD Research Student | 21 | 27 |
| Research Support / Technician | 8 | 10 |
| Other | 19 | 24 |
| **Totals** | **79** | **100** |

308

309

310

311 *Table 4: Results of Howard CSE measure based on 7-point Liker scale where 1 = 'Strongly disagree' and 7 = 'Strongly*
312 *agree'. Mean, median and standard deviations presented across all participants and segmented by ASJC area.*

| CSE statements | All participants | | | Life Sciences | | | Physical Sciences | | | Social Sciences | | |
|---|---|---|---|---|---|---|---|---|---|---|---|---|
| | *M* | *Mdn* | *SD* | *M* | *Mdn* | *SD (IQR)* | *M* | *Mdn* | *SD (IQR)* | *M* | *Mdn* | *SD (IQR)* |
| *I can always manage to solve difficult computer problems if I try hard enough.* | 4.90 | 5 | 1.52 | 4.33 | 5 | 1.87 | 5.10 | 5.5 | 1.65 | 4.92 | 5 | 1.26 |
| *If my computer is 'acting-up', I can find a way to get what I want.* | 5.24 | 5.5 | 1.50 | 4.91 | 5 | 1.92 | 5.50 | 6 | 1.43 | 5.14 | 5 | 1.42 |
| *It is easy for me to accomplish my computer goals.* | 5.32 | 6 | 1.30 | 5.17 | 6 | 1.70 | 5.50 | 5.5 | 1.17 | 5.22 | 6 | 1.27 |
| *I am confident that I could deal efficiently with unexpected computer events.* | 4.92 | 5 | 1.53 | 4.42 | 5 | 1.98 | 5.23 | 5 | 1.52 | 4.84 | 5 | 1.36 |
| *I can solve most computer programs if I invest the necessary effort.* | 5.11 | 6 | 1.58 | 4.75 | 5.5 | 2.18 | 5.47 | 6 | 1.33 | 4.95 | 5 | 1.53 |
| *I can remain calm when facing computer difficulties because I can rely on my abilities.* | 4.68 | 5 | 1.66 | 4.42 | 4.5 | 1.78 | 5.10 | 5 | 1.67 | 4.42 | 5 | 1.59 |
| *When I am confronted with a computer problem, I can usually find several solutions.* | 4.86 | 5 | 1.47 | 5.42 | 5.5 | 1.44 | 4.93 | 5 | 1.46 | 4.61 | 5 | 1.48 |
| *I can usually handle whatever computer problem comes my way.* | 5.00 | 5 | 1.40 | 5.17 | 5 | 1.47 | 5.28 | 5 | 1.44 | 4.72 | 5 | 1.32 |
| *Failing to do something on the computer makes me try harder.* | 4.72 | 5 | 1.50 | 4.50 | 5 | 1.78 | 4.93 | 5 | 1.53 | 4.61 | 5 | 1.40 |
| *I am a self-reliant person when it comes to doing things on a computer.* | 5.31 | 5 | 1.56 | 5.42 | 6 | 1.51 | 5.47 | 6 | 1.76 | 5.14 | 5 | 1.42 |
| *There are few things that I cannot do on a computer.* | 4.57 | 5 | 1.82 | 4.00 | 5 | 2.41 | 4.43 | 4.5 | 1.94 | 4.89 | 5 | 1.43 |
| *I can persist and complete most any computer-related task* | 5.08 | 5 | 1.46 | 5.58 | 5.5 | 1.16 | 5.30 | 5.5 | 1.51 | 4.72 | 5 | 1.45 |
| **CSE measure across participants** | **4.98** | **5** | **1.52** | **4.84** | **5** | **1.77** | **5.19** | **5.25** | **1.53** | **4.85** | **5** | **1.41** |
| **Total CSE score across participants (max. = 84)** | **59.09** | **60** | **14.60** | **57.67** | **63.50** | **17.09 (20.50)** | **62.07** | **61.50** | **15.19 (15.50)** | **58.24** | **60** | **12.59 (15.25)** |

313

314

315
316

*Table 5: Results of CSE measure segmented by participant job role. Mean, median and standard deviations presented across job roles for scores per CSE measure and total CSE scores across participants (where possible max. = 84).*

| Academic job role | CSE scores per measure | | | Total CSE scores across participants | | | |
|---|---|---|---|---|---|---|---|
| | *M* | *Mdn* | *SD* | *M* | *Mdn* | *SD* | *IQR* |
| **Lecturer** | 4.56 | 4.96 | 1.58 | 47.67 | 50.50 | 19.09 | 27.25 |
| **Professor / Reader** | 4.60 | 4.79 | 1.12 | 54.30 | 57.50 | 13.28 | 16.75 |
| **PhD Student** | 5.21 | 5.33 | 1.14 | 62.48 | 64.00 | 13.64 | 17 |
| **Postdoc** | 5.65 | 5.92 | 0.48 | 67.80 | 71.00 | 5.81 | 4 |
| **Research Assistant** | 6.25 | 6.50 | 0.82 | 75.00 | 78.00 | 9.85 | 9.5 |
| **Research Fellow** | 5.33 | 5.00 | 0.87 | 64.00 | 60.00 | 10.42 | 10 |
| **Research Support** | 4.91 | 4.79 | 0.95 | 58.50 | 57.50 | 12.20 | 19 |
| **Other** | 4.59 | 4.75 | 1.33 | 55.11 | 57.00 | 16.01 | 21.5 |

317
318

319

320 *Table 6: Results of the four PID recognition tests from section 2 of the research instrument. Responses provided as %, with n in parentheses. Correct responses to the challenge are denoted by*
321 *an asterisk (\*).*

| Available responses | Task #1 | Task #2 | Task #3 | Task #4 |
|---|---|---|---|---|
| *Publications* | 84.81 (67)* | 78.48 (62)* | 75.95 (60)* | 86.08 (68)* |
| *Publication on repository* | 50.63 (40) | 65.82 (52)* | 41.77 (33) | 60.76 (48)* |
| *Research data* | 8.86 (7) | 6.33 (5) | 74.68 (59)* | 7.59 (6) |
| *Research grants* | 5.06 (4) | 0 (0) | 1.27 (1) | 0 (0) |
| *Organizations* | 24.05 (19) | 0 (0) | 8.86 (7) | 3.80 (3) |
| *Software* | 0 (0) | 0 (0) | 2.53 (2) | 0 (0) |
| *People* | 84.81 (67)* | 5.06 (4) | 17.72 (14) | 2.53 (2) |
| *Instruments* | 1.27 (1) | 0 (0) | 3.80 (3) | 1.27 (1) |
| *Equipment* | 0 (0) | 0 (0) | 1.27 (1) | 0 (0) |
| *Projects* | 7.59 (6) | 2.53 (2) | 2.53 (2) | 1.27 (1) |
| *Audiovisual* | 1.27 (1) | 0 (0) | 0 (0) | 0 (0) |
| *Metadata* | 35.44 (28) | 12.66 (10) | 30.38 (24) | 7.59 (6) |
| *None* | 0 (0) | 1.27 (1) | 1.27 (1) | 5.06 (4) |

322

323

324
325

*Table 7: Results of the four PID recognition tests from section 2 of the research instrument by participant discipline. Responses provided as %. Correct responses to the challenge are denoted by an asterisk (\*).*

| Available responses | Physical Sciences | | | | Social Sciences | | | | Life Sciences | | | |
|---|---|---|---|---|---|---|---|---|---|---|---|---|
| | Task #1 | Task #2 | Task #3 | Task #4 | Task #1 | Task #2 | Task #3 | Task #4 | Task #1 | Task #2 | Task #3 | Task #4 |
| *Publications* | 86.67* | 83.33* | 76.67* | 90.00* | 78.38* | 70.27* | 70.27* | 78.38* | 78.38* | 91.67* | 91.67* | 100.00* |
| *Publication on repository* | 60.00 | 56.67* | 50.00 | 53.33* | 43.24 | 67.57* | 35.14 | 56.76* | 43.24 | 83.33* | 41.67 | 91.67* |
| *Research data* | 3.33 | 6.67 | 86.678 | 6.67 | 13.51 | 8.11 | 62.16* | 10.81 | 13.51 | 0.00 | 91.67* | 0.00 |
| *Research grants* | 6.67 | 0.00 | 0.00 | 0.00 | 5.41 | 0.00 | 2.70 | 0.00 | 5.41 | 0.00 | 0.00 | 0.00 |
| *Organizations* | 26.67 | 0.00 | 13.33 | 3.33 | 27.03 | 0.00 | 8.11 | 5.41 | 27.03 | 0.00 | 0.00 | 0.00 |
| *Software* | 0.00 | 0.00 | 0.00 | 0.00 | 0.00 | 0.00 | 5.41 | 0.00 | 0.00 | 0.00 | 0.00 | 0.00 |
| *People* | 93.33* | 6.67 | 23.33 | 0.00 | 75.68* | 5.41 | 16.22 | 5.41 | 75.68* | 0.00 | 8.33 | 0.00 |
| *Instruments* | 0.00 | 0.00 | 3.33 | 0.00 | 2.70 | 0.00 | 5.41 | 2.70 | 2.70 | 0.00 | 0.00 | 0.00 |
| *Equipment* | 0.00 | 0.00 | 0.00 | 0.00 | 0.00 | 0.00 | 2.70 | 0.00 | 0.00 | 0.00 | 0.00 | 0.00 |
| *Projects* | 10.00 | 3.33 | 0.00 | 0.00 | 5.41 | 2.70 | 5.41 | 2.70 | 5.41 | 0.00 | 0.00 | 0.00 |
| *Audiovisual* | 3.33 | 0.00 | 0.00 | 0.00 | 0.00 | 0.00 | 0.00 | 0.00 | 0.00 | 0.00 | 0.00 | 0.00 |
| *Metadata* | 40.00 | 10.00 | 33.33 | 10.00 | 35.14 | 13.51 | 32.43 | 5.41 | 35.14 | 16.67 | 16.67 | 8.33 |
| *None* | 0.00 | 3.33 | 3.33 | 6.67 | 0.00 | 0.00 | 0.00 | 5.41 | 0.00 | 0.00 | 0.00 | 0.00 |

326
327

328
329 *Table 8: Aggregated results of the four PID recognition tests from section 2 of the research instrument by discipline, including total aggregated score attained by discipline and by job role, and its percentage equivalent. Measures of central tendency provided for individual performance on recognition tests by discipline.*

| Discipline | Total possible score | Total score attained | Total score as % of possible | Mean (participant score) | Median (participant score) | SD (participant score) | IQR (participant score) |
|---|---|---|---|---|---|---|---|
| *Life Sciences* | 96 | 82 | 85.42 | 6.83 | 7.00 | 1.40 | 1.25 |
| *Physical Sciences* | 240 | 168 | 70.00 | 5.60 | 6.00 | 1.99 | 2.75 |
| *Social Sciences* | 296 | 194 | 65.54 | 5.24 | 6.00 | 2.89 | 5.00 |
| **Job role** | | | | | | | |
| *Prof* | 57 | 80 | 71.25 | 5.70 | 7.00 | 3.02 | 3.50 |
| *Lecturer* | 33 | 48 | 68.75 | 5.50 | 6.50 | 3.08 | 3.25 |
| *Research Fellow* | 28 | 56 | 50.00 | 4.00 | 5.00 | 3.00 | 3.50 |
| *Research Assistant* | 23 | 24 | 95.83 | 7.67 | 8.00 | 0.58 | 0.50 |
| *Postdoc* | 33 | 40 | 82.50 | 6.60 | 7.00 | 1.14 | 1.00 |
| *PhD* | 122 | 168 | 72.62 | 5.81 | 6.00 | 1.57 | 2.00 |
| *Research Support* | 41 | 64 | 64.06 | 5.13 | 6.50 | 3.31 | 5.50 |
| *Other* | 107 | 152 | 70.39 | 5.63 | 6.00 | 2.43 | 3.00 |

330
331
332
333
334
335

336

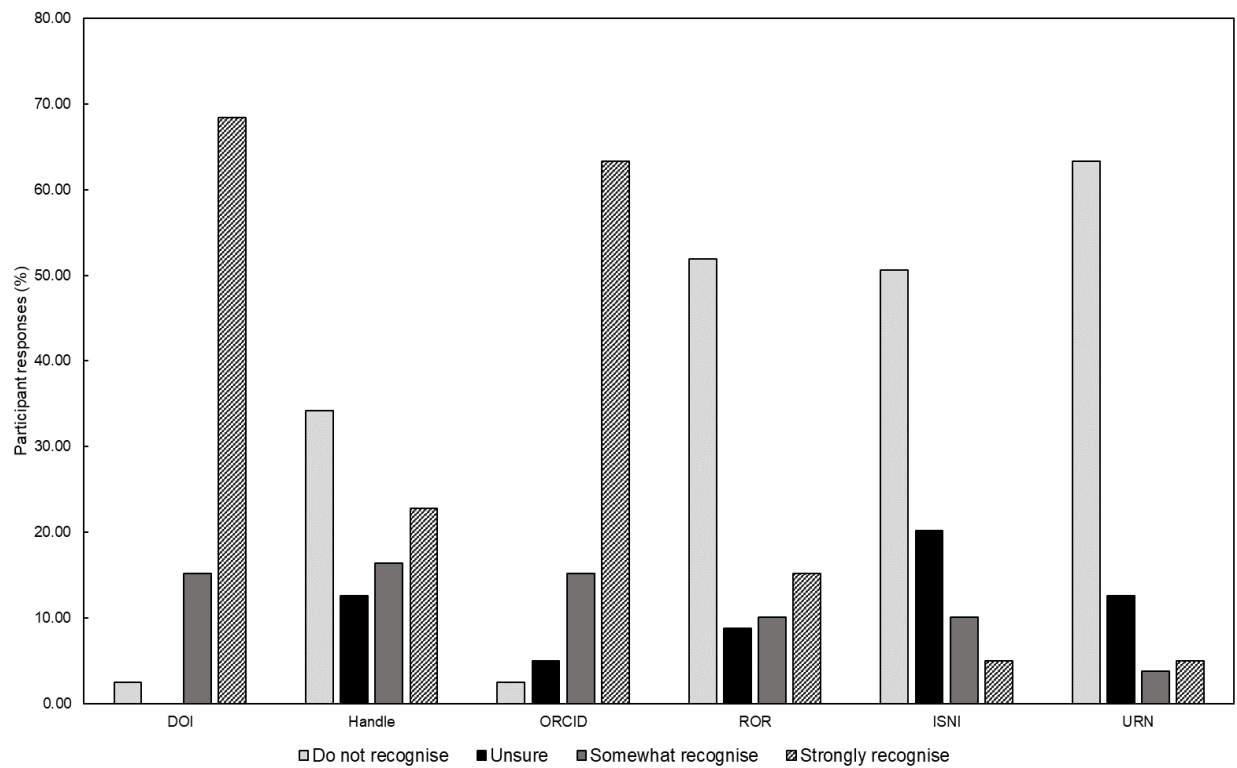

337

338

*Figure 5: Recognition levels of specific PID schemes by participants. Participant responses reported as percentage of total participants.*

339

*Table 9: Entity types participants most associated with the PID schemes presented as part of the PID recognition tests.*

| Entity type | DOI | Handle | ORCID | ROR | ISNI | URN |
|---|---|---|---|---|---|---|
| Publications (on publisher website or platform) | 77.22 | 12.66 | 12.66 | 2.53 | 5.06 | 7.59 |
| Publications (on repository) | 36.71 | 31.65 | 13.92 | 1.27 | 3.80 | 5.06 |
| Research data or open data | 18.99 | 13.92 | 5.06 | 2.53 | 2.53 | 6.33 |
| Research grants | 1.27 | 0.00 | 0.00 | 1.27 | 0.00 | 0.00 |
| Organizations | 0.00 | 0.00 | 3.80 | 21.52 | 6.33 | 1.27 |
| Software | 2.53 | 2.53 | 0.00 | 0.00 | 0.00 | 2.53 |
| People (e.g. authors, editors, Pis, etc.) | 0.00 | 0.00 | 68.35 | 0.00 | 6.33 | 1.27 |
| Research instruments | 1.27 | 0.00 | 0.00 | 0.00 | 0.00 | 0.00 |
| Research equipment or facilities | 1.27 | 0.00 | 0.00 | 0.00 | 0.00 | 0.00 |
| Projects or research activities | 2.53 | 2.53 | 1.27 | 1.27 | 0.00 | 0.00 |
| Audiovisual resources | 3.80 | 3.80 | 0.00 | 0.00 | 0.00 | 3.80 |
| Metadata (bibliographic data) | 6.33 | 2.53 | 6.33 | 0.00 | 2.53 | 0.00 |
| None | 2.53 | 1.27 | 2.53 | 1.27 | 2.53 | 3.80 |

348

349 *Table 10: Factor scores of ratings of PID concepts, segmented by semantic dimension and by combined concept score (CCS). Concept tested against the bipolar adjectives presented in Table 1.*

| Concept tested | All participants | | | |
|---|---|---|---|---|
| | **Evaluation** | **Potency** | **Activity** | **CSS** |
| *Scholarly communications* | 1.84 | 2.09 | 0.55 | 1.49 |
| *People* | 1.78 | 1.87 | 0.72 | 1.46 |
| *Places* | 1.13 | 1.04 | 0.21 | 0.79 |
| *Things* | 1.63 | 1.81 | 0.13 | 1.19 |
| **Factor mean** | **1.60** | **1.70** | **0.40** | **1.23** |

350

351

352

353

354 *Table 10: Factor scores of ratings of PID concepts, segmented by semantic dimension, combined concept score (CCS) and by discipline. Concept tested against the bipolar adjectives presented*
355 *in Table 1.*

| Concept tested | Physical Sciences | | | | Social Sciences | | | | Life Sciences | | | |
|---|---|---|---|---|---|---|---|---|---|---|---|---|
| | **Evaluation** | **Potency** | **Activity** | **CCS*** | **Evaluation** | **Potency** | **Activity** | **CCS** | **Evaluation** | **Potency** | **Activity** | **CCS** |
| *Scholarly communications* | 1.71 | 1.78 | 0.39 | 1.29 | 1.84 | 2.17 | 0.84 | 1.62 | 2.33 | 2.25 | 0.42 | 1.67 |
| *People* | 1.24 | 1.19 | 0.35 | 0.93 | 1.83 | 2.00 | 0.60 | 1.48 | 3.61 | 3.31 | 3.08 | 3.33 |
| *Places* | 0.83 | 0.56 | 0.32 | 0.57 | 1.30 | 1.30 | 0.14 | 0.91 | 0.83 | 0.83 | -0.56 | 0.37 |
| *Things* | 1.31 | 1.44 | -0.19 | 0.85 | 1.63 | 1.84 | 0.21 | 1.23 | 2.55 | 2.63 | 0.75 | 1.98 |
| **Factor mean** | **1.27** | **1.24** | **0.22** | **0.91** | **1.65** | **1.83** | **0.45** | **1.31** | **2.33** | **2.26** | **0.92** | **1.84** |

356 * CcS = 'combined concept score' by mean across all dimensions (evaluation, potency, activity) for a test concept.

357

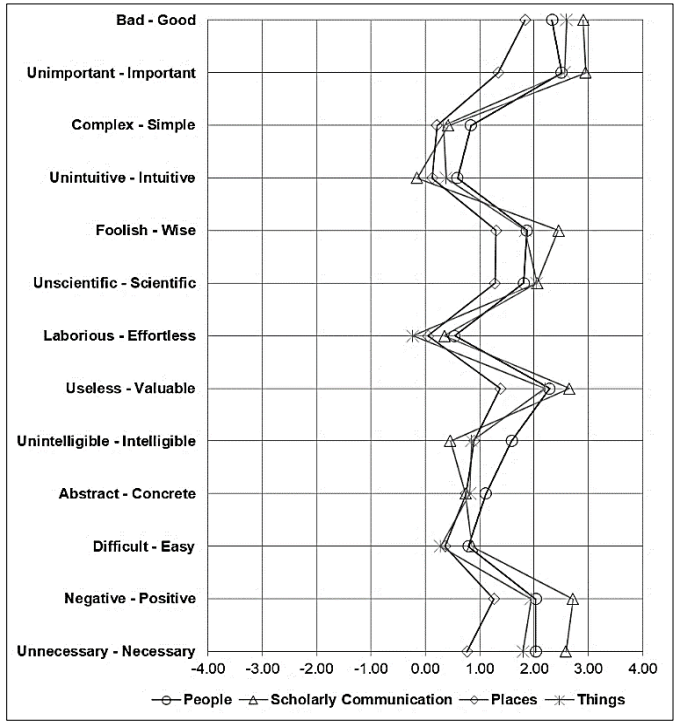

**A.**

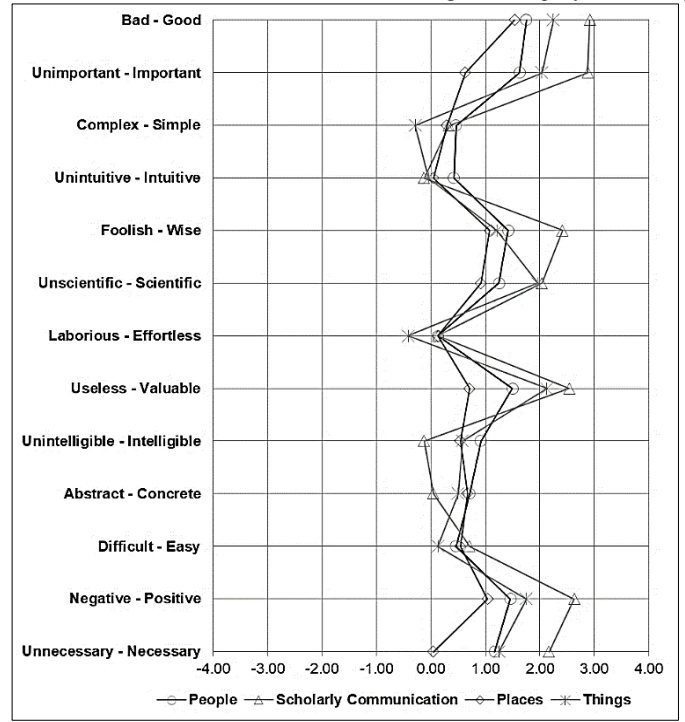

**B.**

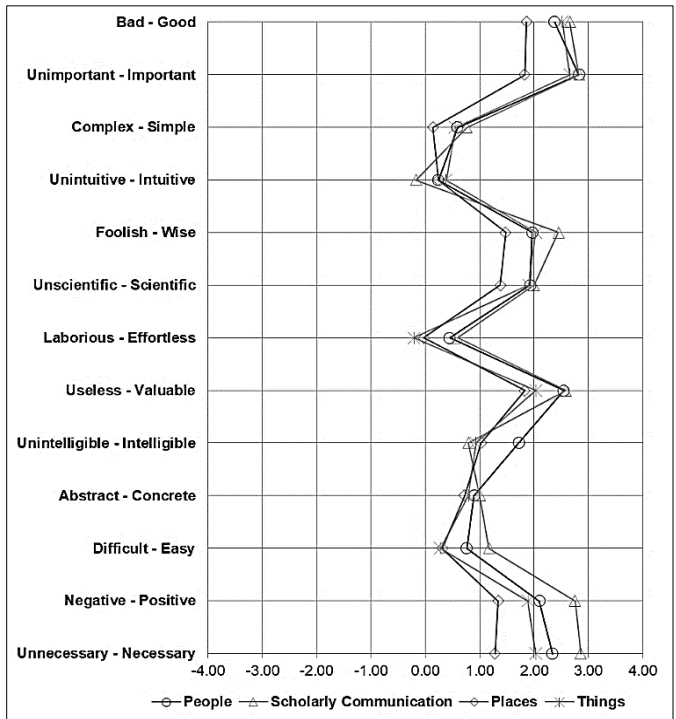

**C.**

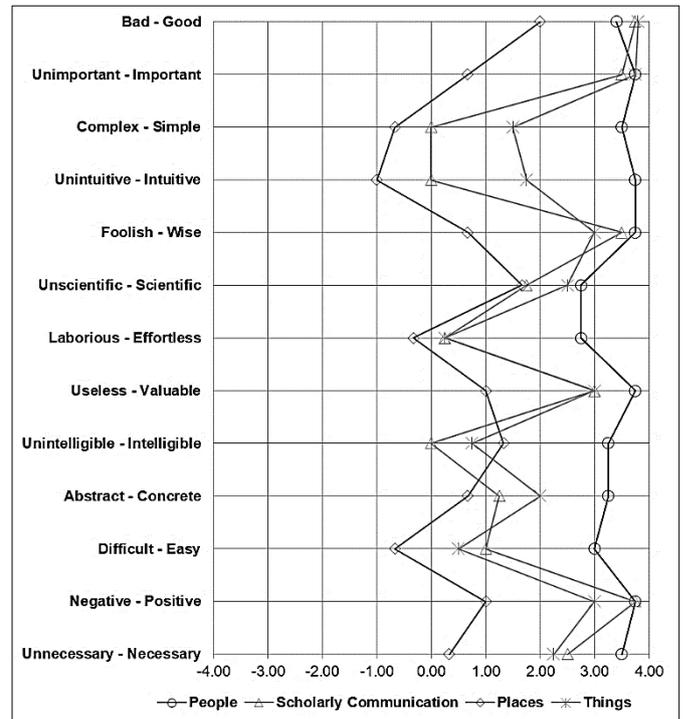

**D.**

*Figure 6: PID perception measurements by all participants (A) and by discipline groupings (B = Physical Sciences; C = Social Sciences; D = Life Sciences), charted using the 'semantic distance notion' approach. Charts plot the tested PID concepts according to their perception on the 13 bipolar adjective pairs comprising the semantic differential scale. Table 1 denotes how these adjective pairs were assigned to semantic dimensions (factors).*

363

364

365 *Table 12: Results of PID perception measurements using semantic differential scale method and data forming basis of semantic distance charts in Figure 6. Data are arranged by semantic scale item and semantic*
366 *space dimension, and by participant discipline grouping semantic distance. Scores on each scale item are the calculated means.*

| Semantic scale item – bi-polar adjective pairs | Semantic space dimension (factor) | Physical Sciences | | | | Social Sciences | | | | Life Sciences | | | |
|---|---|---|---|---|---|---|---|---|---|---|---|---|---|
| | | Scholarly communications | People | Places | Things | Scholarly communications | People | Places | Things | Scholarly communications | People | Places | Things |
| Bad - Good | *Evaluation* | 2.92 | 1.75 | 1.54 | 2.25 | 2.66 | 2.38 | 1.86 | 2.52 | 3.75 | 3.40 | 2.00 | 3.80 |
| Unimportant - Important | *Potency* | 2.88 | 1.63 | 0.63 | 2.04 | 2.83 | 2.83 | 1.83 | 2.66 | 3.50 | 3.75 | 0.67 | 3.75 |
| Complex - Simple | *Activity* | 0.33 | 0.46 | 0.29 | -0.29 | 0.76 | 0.59 | 0.14 | 0.55 | 0.00 | 3.50 | -0.67 | 1.50 |
| Unintuitive - Intuitive | *Evaluation* | -0.13 | 0.42 | 0.04 | -0.04 | -0.17 | 0.24 | 0.24 | 0.38 | 0.00 | 3.75 | -1.00 | 1.75 |
| Foolish - Wise | *Evaluation* | 2.42 | 1.42 | 1.08 | 1.21 | 2.45 | 1.97 | 1.48 | 2.03 | 3.50 | 3.75 | 0.67 | 3.00 |
| Unscientific - Scientific | *Potency* | 2.04 | 1.25 | 0.92 | 1.96 | 2.00 | 1.93 | 1.38 | 1.90 | 1.75 | 2.75 | 1.67 | 2.50 |
| Laborious - Effortless | *Activity* | 0.13 | 0.13 | 0.13 | -0.42 | 0.59 | 0.45 | -0.03 | -0.21 | 0.25 | 2.75 | -0.33 | 0.25 |
| Useless - Valuable | *Evaluation* | 2.54 | 1.50 | 0.71 | 2.13 | 2.59 | 2.55 | 1.83 | 2.03 | 3.00 | 3.75 | 1.00 | 3.00 |
| Unintelligible - Intelligible | *Evaluation* | -0.13 | 0.92 | 0.54 | 0.58 | 0.79 | 1.72 | 1.03 | 0.93 | 0.00 | 3.25 | 1.33 | 0.75 |
| Abstract - Concrete | *Potency* | 0.04 | 0.71 | 0.67 | 0.50 | 1.00 | 0.90 | 0.72 | 0.79 | 1.25 | 3.25 | 0.67 | 2.00 |
| Difficult - Easy | *Activity* | 0.71 | 0.46 | 0.54 | 0.13 | 1.17 | 0.76 | 0.31 | 0.28 | 1.00 | 3.00 | -0.67 | 0.50 |
| Negative - Positive | *Evaluation* | 2.63 | 1.46 | 1.04 | 1.75 | 2.76 | 2.10 | 1.34 | 1.89 | 3.75 | 3.75 | 1.00 | 3.00 |
| Unnecessary - Necessary | *Potency* | 2.17 | 1.17 | 0.04 | 1.25 | 2.86 | 2.34 | 1.28 | 2.03 | 2.50 | 3.50 | 0.33 | 2.25 |

367

368

369

370

371

372      *Table 13: Semantic distance measures (D) relating to the PID concepts tested, by all participants and by discipline.*

| PID Concept | Scholarly comms. | People | Places | Things |
|---|---|---|---|---|
| *All participants* | | | | |
| *Scholarly communications* | | | | |
| *People* | 0.29 | | | |
| *Places* | 1.31 | 2.04 | | |
| *Things* | 1.31 | 0.62 | 0.92 | |
| *Physical Sciences* | | | | |
| *Scholarly communications* | | | | |
| *People* | 0.76 | | | |
| *Places* | 1.51 | 1.26 | | |
| *Things* | 1.17 | 0.60 | 1.13 | |
| *Social Sciences* | | | | |
| *Scholarly communications* | | | | |
| *People* | 0.30 | | | |
| *Places* | 1.24 | 2.12 | | |
| *Things* | 1.08 | 0.46 | 0.64 | |
| *Life Sciences* | | | | |
| *Scholarly communications* | | | | |
| *People* | 3.14 | | | |
| *Places* | 2.28 | 5.65 | | |
| *Things* | 2.25 | 2.65 | 2.80 | |

373

374

375

376   Table 14: Results of Wilcoxon signed-rank test on semantic distance (D) between PID concepts by discipline grouping. P-values at < .05
377                      indicated by asterisk (*); p-values at .01 indicated by double asterisk (**).

| Participant discipline groupings tested | Semantic distance tested | T | z | p-value |
|---|---|---|---|---|
| Physical Sciences / Life Sciences | | 3.58 | 2.11 | 0.02* |
| Physical Sciences / Social Sciences | Scholarly communications / Things | 0.76 | 0.32 | 0.37 |
| Social Sciences / Life Sciences | | 3.74 | 2.19 | 0.01** |
| Physical Sciences / Life Sciences | | 1.46 | 1.36 | 0.09 |
| Physical Sciences / Social Sciences | People / Places | 0.82 | 0.51 | 0.31 |
| Social Sciences / Life Sciences | | 1.15 | 1.26 | 0.10 |
| Physical Sciences / Life Sciences | | 0.58 | 0.72 | 0.24 |
| Physical Sciences / Social Sciences | Scholarly communications / Places | 0.28 | 0.32 | 0.37 |
| Social Sciences / Life Sciences | | 0.42 | 0.58 | 0.28 |
| Physical Sciences / Life Sciences | | 0.14 | 0.03 | 0.49 |
| Physical Sciences / Social Sciences | People / Things | 0.96 | 0.27 | 0.39 |
| Social Sciences / Life Sciences | | 0.45 | 0.19 | 0.42 |
| Physical Sciences / Life Sciences | | 0.22 | 0.20 | 0.42 |
| Physical Sciences / Social Sciences | Places / Things | 2.52 | 2.56 | 0.01** |
| Social Sciences / Life Sciences | | 1.34 | 1.36 | 0.09 |
| Physical Sciences / Life Sciences | | 0.77 | 0.38 | 0.35 |
| Physical Sciences / Social Sciences | Scholarly communications / People | 0.09 | 0.53 | 0.30 |
| Social Sciences / Life Sciences | | 0.78 | 0.92 | 0.18 |

378

379   Table 15: Participants' reported familiarity of using PIDs in scholarly communication and knowledge of the purpose of PIDs, by all participants
380                                          and by discipline.

| | All participants | | Physical Sciences | | Social Sciences | | Life Sciences | |
|---|---|---|---|---|---|---|---|---|
| | *(Un)familiar* | *(Un)knowledgable* | *(Un)familiar* | *(Un)knowledgable* | *(Un)familiar* | *(Un)knowledgable* | *(Un)familiar* | *(Un)knowledgable* |
| *M* | 7.68 | 7.29 | 7.00 | 6.60 | 8.14 | 7.86 | 7.60 | 6.80 |
| *Mdn* | 9.00 | 8.00 | 8.00 | 7.00 | 9.00 | 8.00 | 9.00 | 9.00 |
| *SD* | 1.81 | 1.89 | 1.64 | 1.52 | 1.77 | 1.68 | 1.97 | 1.91 |

381

382   Table 16: Participants' reported familiarity of using PIDs in scholarly communication and knowledge of the purpose of PIDs, by job role.

| *(Un)familiar* | | | | | | | |
|---|---|---|---|---|---|---|---|
| | Professor / Reader | Lecturer | Research Fellow | Research Assistant | Postdoc | PhD Research Student | Research Support | Other |
| *M* | 8.10 | 8.00 | 8.43 | 7.00 | 8.60 | 6.48 | 8.75 | 8.67 |
| *Mdn* | 9.00 | 9.00 | 9.00 | 7.00 | 9.00 | 7.00 | 9.00 | 9.00 |
| *SD* | 0.41 | 0.81 | 1.22 | 1.63 | 2.03 | 2.44 | 2.84 | 3.25 |
| *(Un)knowledgable* | | | | | | | |
| *M* | 8.00 | 6.60 | 8.00 | 6.00 | 8.40 | 6.24 | 8.25 | 8.67 |
| *Mdn* | 9.00 | 6.00 | 9.00 | 5.00 | 9.00 | 6.00 | 8.00 | 9.00 |
| *SD* | 0.41 | 0.81 | 1.22 | 1.63 | 2.03 | 2.44 | 2.84 | 3.25 |

383

384

385

386

387

388

*Table 17: Participant views on the purpose of PIDs and percentage of participants indicating agreement with purpose.*

| Purpose of PIDs | Participant response to stated PID purpose (%) |
|---|---|
| To protect against links (URLs or 'web addresses') that may become broken over time (i.e. 'link rot') | 78.26 |
| To ensure the persistent and unambiguous citation of scholarly objects on the web | 81.16 |
| To promote interlinking between scholarly objects on the web | 68.12 |
| To promote the findability of my scholarly work | 56.52 |
| To ensure long-term maintenance and integrity of the published scholarly record on the web (e.g. for the purposes of verification, reanalysis, study reproduction, replication, etc.) | 73.91 |
| To enrich global bibliographic data about scholarly objects on the web and beyond | 49.28 |
| To support more accurate counting and tracking of citations of my work and the work of others | 56.52 |
| To assist in the tracking of the alternative impact of scholarly objects | 39.13 |

389

390

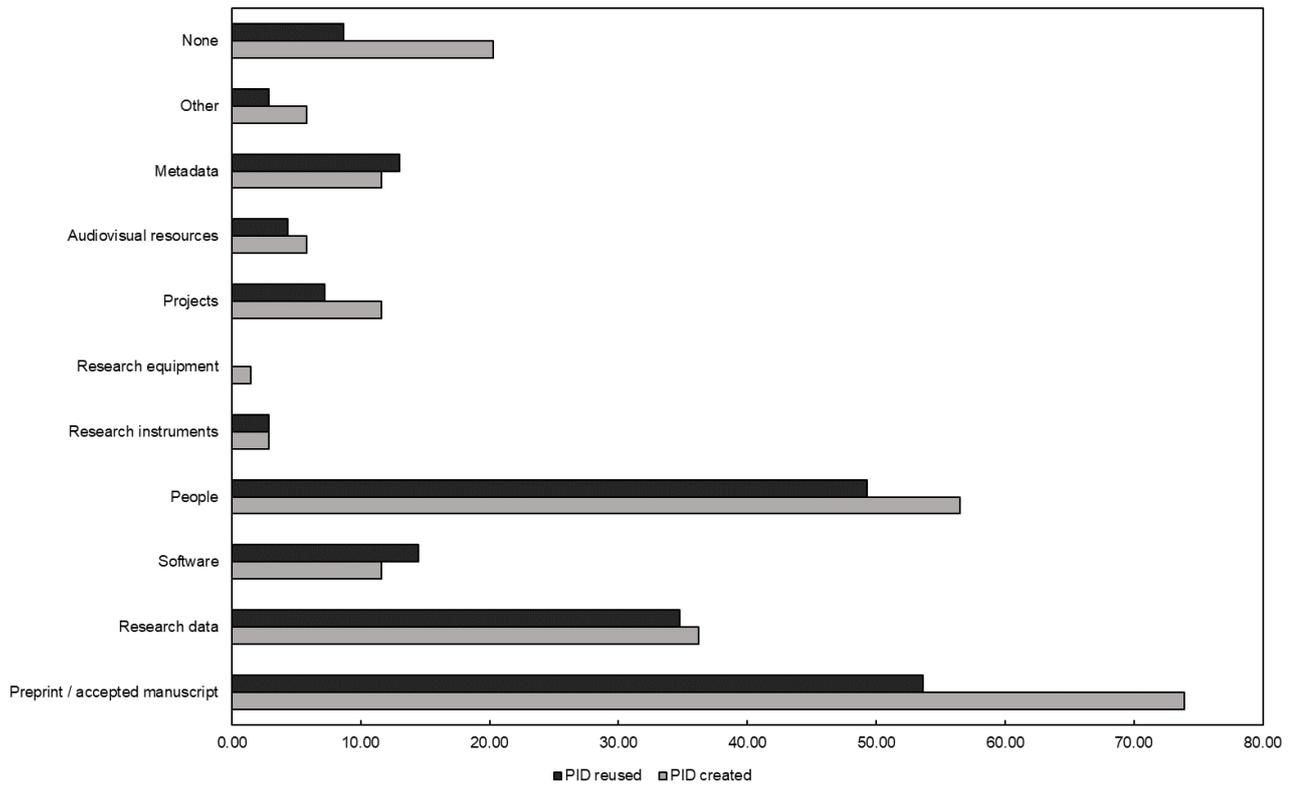

394    *Figure 7: Participant reporting on entities for which PIDs had been created and (re)used over the past 4 years.*